\documentclass{aastex631}

\shorttitle{The Distribution of Highly Red-Sloped Asteroids in the Middle and Outer Main Belt}
\shortauthors{Humes et al.}

\graphicspath{{./}{figures/}}

\begin{document}

\title{The Distribution of Highly Red-Sloped Asteroids in the Middle and Outer Main Belt}

\correspondingauthor{Oriel Humes}
\email{oah28@nau.edu}

\author[0000-0002-1700-5364]{Oriel A. Humes}
\affiliation{Northern Arizona University \\
Flagstaff, Arizona \\
86011, USA}

\author[0000-0003-3091-5757]{Cristina A. Thomas}
\affiliation{Northern Arizona University \\
Flagstaff, Arizona \\
86011, USA}

\author{Lauren E. McGraw}
\affiliation{Northern Arizona University \\
Flagstaff, Arizona \\
86011, USA}

\begin{abstract}

Red ($S > 10 \% /0.1 \mu$m) spectral slopes are common among Centaurs and Trans-Neptunian Objects (TNOs) in the outer Solar System. Interior to and co-orbital with Jupiter, the red ($S \sim 10 \% /0.1 \mu$m) slopes of D-type Main Belt and Jupiter Trojan asteroids are thought to reflect their hypothesized shared origin with TNOs beyond the orbit of Jupiter. In order to quantify the abundance of red-sloped asteroids within the Main Belt, we conducted a survey using the NASA Infrared Telescope Facility and the Lowell Discovery Telescope. We followed up on 32 candidate red objects identified via spectrophotometry from the Sloan Digital Sky Survey's Moving Object Catalog to confirm their steep spectral slopes and determine their taxonomic classifications. We find that our criteria for identifying candidate red objects from the Moving Object Catalog result in a $\sim 50\%$ confirmation rate for steeply red sloped asteroids. We also compare our observations of Main Belt asteroids to existing literature spectra of the Jupiter Trojans and steeply red-sloped Main Belt asteroids. We show that some red sloped asteroids have linearly increasing reflectance with increasing wavelength, while other red-sloped asteroids show a flattening in slope at longer near-infrared wavelengths, indicating a diversity among the population of spectrally red Main Belt asteroids suggestive of a variety of origins among the population of steep sloped asteroids.

\end{abstract}


\section{Introduction} \label{sec:intro}

Many outer Solar System populations, including trans-Neptunian Objects (TNOs), comets, and Centaurs are characterized by their red ($S > 10 \% /0.1 \mu$m) colors in the visible and near infrared (VNIR) (e.g. \cite{sheppard2010colors, perna2010colors, lamy2009colors}). That is, their VNIR spectra show increasing reflectance with increasing wavelength. To precisely quantify the `redness' of a spectrum, many studies use spectral slope ($S$), typically measured as a percentage increase in reflectance over a certain spectral range, usually 0.1 $\mu$m, relative to a reference wavelength, typically 0.5 $\mu$m (\cite{hainaut2012colours, doressoundiram2008color}). The colors of small bodies can also be quantified using color indices, or the difference in astronomical magnitude of an object as seen in different standard filter sets, a photometric technique that is particularly useful for quantifying the color of distant, faint objects like TNOs. Among TNOs, the reddest surfaces are found among the cold Classical Kuiper Belt object population and the extremely distant inner Oort cloud objects which are thought to have formed and remained far from the Sun (\cite{sheppard2010colors, doressoundiram2008color}). Some authors have interpreted this trend to reflect a primordial color gradient within the early Solar nebula, with objects that initially formed at high heliocentric distances retaining more volatile organics, which become red upon irradiation (\cite{brown2011hypothesis}).

The general trend of increasing abundance of spectrally red materials at increasing distances is also seen among small body populations interior to and co-orbital with Jupiter. In the Main Belt, the proportion of dark, red-slope asteroids increases with increasing heliocentric distance (\cite{gradie1982, demeo2014}). Further out, in the Jupiter Trojans, a population of asteroids co-orbital with Jupiter at 5.2 AU, asteroids are classified into the less red (LR) and red (R) spectral groups, with the majority of Trojans belonging to the R group \cite{Emery_2010}. The average VNIR slope of an R group Trojan ($\sim 10\% /0.1 \mu$m, \cite{Wong_2014}) is slightly steeper than a typical D-type Main Belt asteroid ($\sim 9\% /0.1 \mu$m, \cite{gartrelle_2021}), the reddest class of asteroids commonly found in the Main Belt. The spectral slopes of small bodies interior to and co-orbital with Jupiter, such as Trojans and D-type Main Belt asteroids, are generally shallower than those of the reddest small bodies in the outer Solar System, such as TNOs and Centaurs, which can have ultra-red slopes as steep as $\sim 40 - 55\% /0.1 \mu$m  (\cite{hainaut2012colours, Wong_2016}), further demonstrating the association of steep red spectral slopes with outer Solar System populations. 

The current distribution of small Solar System bodies does not perfectly reflect initial conditions within the protoplanetary nebula (e.g. \cite{bottke2006yarkovsky, levison2009, walsh2012populating, demeo2014}). The processes of planetary formation and migration have led to the displacement of small bodies from their formation locations via gravitational interaction, with many models positing that the migration of the gas giant planets resulted in the delivery of outer Solar System materials from the proto-Kuiper Belt to the Main Belt (\cite{levison2009, walsh2012populating}), resulting in the diversity of asteroid spectral types we see in the Main Belt today. Based on their spectral similarity to TNOs and Centaurs, red-sloped D-types and R group Trojans have been identified as possible migrants that originated from a source population beyond the present day orbit of Jupiter (\cite{demeo2014,Wong_2016}). Dynamical simulations also support this idea, suggesting the Trojans were captured from the same parent population as the present day TNOs (\cite{Morbidelli_2005,Nesvorny_2013,pirani_2019}). Thus, the presence and distribution of red, TNO-like material can be used to provide indirect constraints on the mechanisms of planetary formation. The constraints provided by observational studies can then be used to test the predictions of dynamical models and further refine their accuracy. The recent serendipitous discovery of extremely red, TNO-like asteroids within the Main Belt has sparked discussion of dynamical implications of the presence of objects that could have migrated from 20-30 AU to Main Belt (\cite{hasegawa2021}). 

The distribution of red material within the Main Belt has been the subject of recent observational surveys. In \cite{demeo2014_dtype}, the distribution of D-types in the inner Main Belt (interior to 2.5 AU) is investigated by identifying D-type ``candidate'' asteroids via Sloan Digital Sky Survey (SDSS) photometry and confirming the taxonomy of these asteroids using near-infrared spectroscopy from the NASA Infrared Telescope Facility (IRTF). Those authors observed 13 candidate D-types in the inner Main Belt and confirmed the D-type taxonomy of three of these candidates, resulting in a $\sim 20\%$ confirmation rate (\cite{demeo2014_dtype}). Similarly, in this work, we follow up on candidate red asteroids with slopes in excess of the average R group Trojan identified using SDSS photometry using the NASA IRTF and Lowell Discovery Telescope (LDT). Unlike \cite{demeo2014_dtype}, we seek to identify asteroids that are redder than the average Main Belt D-types, using the slightly steeper average spectral slope of the R group Trojans as a stricter criterion for identifying those steeply-sloped red objects currently located in the Main Belt of potentially sourced from the same population as the TNOs. 

\section{Methods}

\subsection{Target Selection}

To identify candidate red Main Belt asteroids, we used the Sloan Digital Sky Survey (SDSS) Moving Object Catalog 4 (MOC) (\cite{ivezic2002asteroids}). The SDSS captures photometric measurements of objects in the \textit{u}, \textit{g}, \textit{r}, \textit{i}, and \textit{z} filters in sequential order. We removed all observations in the MOC with apparent magnitudes fainter than 22.0, 22.2, 22.2, 21.3, and 20.5 for each of the \textit{u}, \textit{g}, \textit{r}, \textit{i}, and \textit{z} filters, respectively, as these magnitudes correspond to the limiting magnitudes for 95\% survey completeness (\cite{ivezic2001solar}). Following \cite{DemeoCarry_2014}, potentially anomalous observations, including those taken during non-photometric conditions or near detector edges, were removed from the target list by removing those observations flagged with \textit{edge, badsky, peakstooclose, notchecked, binned4, nodeblend, deblenddegenerate, badmovingfit, toofewgooddetections,} and \textit{stationary}. These conditions narrowed the target list from over 220,000 observations to approximately 68,000 moving object observations.

The asteroid magnitudes reported in the MOC were converted to measurements of spectrophotometric reflectance by calculating color indices for each asteroid (\textit{g-r}, \textit{r-i}, and \textit{i-z}), subtracting solar colors from each color index, converting each magnitude difference to a flux ratio, and normalizing to the \textit{r} band reflectance value. The effective wavelength of each observation was estimated as the center wavelengths of the \textit{u}, \textit{g}, \textit{r}, \textit{i}, and \textit{z} filters, (0.3551, 0.4686, 0.6166, 0.7480, and 0.8923 $\mu$m, respectively). Spectral slopes were measured directly from the photometric data by fitting a line to the solar corrected \textit{g}, \textit{r}, \textit{i}, and \textit{z} measurements. To quantify the error in slope, we generated 20,000 synthetic spectrophotometric measurements for each candidate asteroid by drawing values from a Gaussian distribution with a mean equal to the observed normalized color indices and error drawn from the propogated error in reflectance. For each synthetic observation, slope was calculated by fitting a line to the synthetic spectrophotometry. The slope error was estimated using the standard deviation of all 20,000 synthetic spectrophotometric measurements. Asteroids with spectral slopes exceeding 12\% / $0.1 \mu$m were chosen as candidate targets to ensure only the reddest objects were selected (the average slope of R group Trojans was found to be 10.3\% / $0.1 \mu$m using the same slope estimation method using data from \cite{Emery_2010}). 

Targets with semimajor axes less than 2.5 AU were excluded, since D-types in the inner Main Belt have already been surveyed using similar methods in \cite{demeo2014_dtype}. This resulted in a list of 75 candidate objects with at least one slope measurement exceeding the 12\% / $0.1 \mu$m among one or more observations in the MOC. We refined this list further by requiring objects with multiple slope measurements to have consistently steep slopes. Similarly, if independent albedo measurements of an object were available (\cite{Mainzer_NEOWISE}), we required objects on the most restrictive final list to have low ($<0.15$) albedos. In total, 60 of the 75 objects were included in the final candidate list. The remaining 15 objects that did not meet the strictest criteria for inclusion were used as a list of lower priority backup targets.

The targets we observed were ultimately determined by their visibility from the observing sites during the survey period. We prioritized asteroids that reached low ($<1.5$) airmasses for at least one hour a night over at least ten days of each observing semester. Because both the NASA Infrared Telescope Facility and Lowell Discovery Telescope are located at mid-latitudes in the Northern Hemisphere, we favored targets with moderately negative to positive declinations during the observing period. We also prioritized targets that were visible from both observing locations in the same semester. To get the best estimate of spectral slope, our signal-to-noise requirement restricted observations to asteroids with V magnitudes $<$ 18.5 for the NASA IRTF and V magnitudes $<$ 19.0 for the Lowell Discovery Telescope.

\subsection{NASA Infrared Telescope Facility Observations}

We used the SpeX Spectrograph on the NASA Infrared Telescope Facility (IRTF) in prism mode (R $\sim$ 200, \cite{rayner2003spex}) to observe candidate asteroids in the near-infrared (0.7 - 2.5 $\mu$m). The IRTF has a long history of use in asteroid surveys (e.g. \cite{marsset2020}), which we leveraged to obtain as accurate measurements of spectral slopes as possible. Observations of each target were taken at airmasses $<2.0$ with the slit aligned to the parallactic angle to reduce the chance of variation in the spectral slope due to differential atmospheric refraction. For bright targets, we used the GuideDog infrared slit viewer to ensure each target was well aligned in the slit, while for faint targets ($V > 17.5$) we used the MIT Optical Rapid Imaging System (MORIS) (\cite{bus2011moris, gulbis2010moris}) to guide on a visible wavelength image of the target to ensure slit alignment. Each observation used a local standard star and a solar standard star to correct for the solar spectrum. Local standards were chosen by querying the SIMBAD astronomical database for bright (12 - 6 V mag) G dwarf stars within a 7 degree radius of the mean position of the asteroid on each night (\cite{simbad}). Solar standard stars were selected from a list of commonly used solar analogs (Bus, S. J., personal comm.). Observations of each asteroid were bracketed by observations of the local standard, returning to obtain local standard spectra after $\sim$ 60 minutes of integration time on each asteroids. Solar standards were observed once a night at low ($< 1.5$) airmasses. This observing strategy allowed us to produce reflectance spectra for each object by averaging different calibrated versions of that object's spectrum from throughout each night. To account for the sky background emission, spectral measurements of each target, solar, and local standards were taken in A-B pairs. Spectra were extracted from raw FITS files using SpeXtool. SpeXtool performs flat field correction, wavelength registration, A-B pair subtractions, telluric correction, and conversion from absolute fluxes to relative reflectance spectra (\cite{cushing2004spextool}). We observed a total of 15 targets identified in the SDSS MOC dataset with the IRTF. 

The correction from absolute flux (uncorrected for solar color) to reflectance was accomplished using both the local and solar standards: first the asteroid spectrum was divided by the spectrum of the local standard to produce a locally-corrected spectrum. Then, the solar analog spectrum was divided by the average spectrum of the local standard to produce a solar comparison spectrum and confirm the similarity of the local standard to the solar analog spectrum. Then, the locally-corrected spectrum was divided by the solar comparison spectrum. The effect of this correction is to account for both variations in local atmospheric transparency (accomplished using the local standard) as well as correct for solar color (using the solar standard). Reflectance spectra taken using the IRTF were normalized to 1 at 1.0 $\mu$m. Typically, the differences between the solar analog and local standard spectra were confined to regions of telluric absorption and introduced minimal variation in slope. For IRTF spectra heavily affected by differences in telluric absorption between the solar and local standard spectra, we corrected the spectra using only the local standard if the local standard used had a J-H color index (\cite{cutri20032mass}) no more than $\pm$ 0.1 magnitudes away from the solar J-H value (\cite{casagrande2012infrared}). Asteroids for which this correction was performed are indicated in Table \ref{tab:IRTF_obs} by a single star listed in the standard column. The choice to include or exclude the solar standard in the reduction did not affect the slope significantly enough to change the classification of an asteroid from red to not red, or vice versa. This finding is in line with the results of \cite{marsset2020}, which found that slope varies by less than 0.1\% / 0.1 $\mu$m per 0.1 difference in airmass: all our targets were observed at airmasses between 1 and 2. This variation in slope is comparable to the slope variance between G2V and G5V stars (\cite{marsset2020}).

\begin{longdeluxetable}{llllcccl}
\tablecaption{Observation circumstances for asteroids observed with the NASA Infrared Telescope Facility \label{tab:IRTF_obs}}
\tablewidth{0pt}
\tablehead{
& \colhead{Provisional}	 & \colhead{Obs. Start} & 	& \colhead{Mag.}	& \colhead{Airmass}	& \colhead{Exp.}	& \colhead{Standard}	 \\
\colhead{Asteroid} & \colhead{Designation}	 & \colhead{(UTC)} & 	\colhead{RA/Dec}	& \colhead{(V)}	& \colhead{Range}	& \colhead{(min)}	& \colhead{Star(s)}	
}
\startdata
(203) & & & & & & & SA 105-56 \\
Pompeja	& A879 SA	& 03-07-2022 09:10 & 12:37:30 -05:20:58	& 13.1	& 1.51-1.15	& 77	& HD 110029	  \\
\hline
(467) & & & & & & & SA 105-56 \\
Laura & A901 AA & 02-05-2023 13:18 & 12:25:29 -07:34:18 & 16.0 & 1.14-1.13 & 40 & HD 108808  \\
\hline
(1947) \\
Iso-Heikkila	& 1935 EA & 07-16-2021 11:56 & 23:05:31 -20:02:47 & 16.6 & 1.51-1.34	& 16	& HD 220145	 \\
\hline
(3248) & & & & & & & SA 105-56 \\
Farinella	& 1982 FK	& 02-17-2022	10:25	& 12:51:27 -02:48:12 &	16.7 &	1.54-1.24	& 64 & HD 111662 \\
\hline
(5819) & & & & & & & SA 115-271 \\
Lauretta	& 1989  UZ4	& 09-07-2021	09:35	& 00:39:27 -00:29:44 &	16.8 &	1.30-1.10 &	60	& HD 4096	\\
\hline
(7562) & & & & & & &  SA 105-56 \\
Kagiroino-Oka &	1986 WO9 &	05-15-2021	10:52	& 15:13:48 -14:44:44 &	16.8 &	1.43-1.25 &	50 &	HD 130958 \\
\hline
(9934) \\
Caccioppoli	& 1985 UC &	08-06-2022	10:58	& 23:09:21 -31:03:37 &	16.4 &	1.77-1.67 & 32 &	HD 219180	 \\
\hline
(21867)	& & & & & & & SA 112-1333 \\
 & 1999 TQ251	& 07-25-2022	11:23	& 20:21:49 -01:49:12 &	17.9 &	1.26-1.1 &	68	& HD 191595	  \\
 \hline
(21867)	& & & & & & & SA 112-1333 \\
& 1999 TQ251	& 07-28-2022	11:37	& 20:19:51 -02:04:31 &	17.9 &	1.48-1.14 &	60	& HD 191595  \\
\hline
(22110)	& & & & & & & SA 112-1333 \\
& 2000 QR7	& 07-20-2021	09:39	& 18:33:21 -19:06:07 &	17.5 &	1.37-1.30	& 52	& HD 172404	 \\
\hline
(22422) & & & & & & & SA 112-1333\\
Kenmount Hill	& 1995 YO5	& 07-25-2022	10:00	& 20:06:51 -28:13:45 &	18.0 &	1.50 &	40	& HD 189327  \\
\hline
(22422) & & & & & & & SA 112-1333\\
Kenmount Hill	& 1995 YO5	& 07-28-2022	09:55	& 20:04:05 -28:27:58 &	18.0 &	1.52-1.50 &	52	& HD 189327  \\
\hline 
(23690)	& & & & & & & SA 105-56 \\
& 1997 JD14	& 05-15-2021	06:00 &	12:26:33 -04:32:01 &	18.1 &	1.16-1.10	& 48 &	HD 108808 	 \\
\hline
(25835) \\
Tomzega & 2000 EO20 & 02-05-2023 14:23 & 11:38:43 -12:15:02 & 18.3 & 1.77-1.27 & 72 & HD 100044  \\
\hline
(25835) \\
Tomzega & 2000 EO20 & 02-10-2023 13:47  &  11:37:07 -12:06:01 & 18.2 & 1.94-1.24 & 92 & HD 100044  \\
\hline
(27378)	& & & & & & & SA 102-1081 \\
& 2000 EG55	& 02-23-2022	09.25 &	10:33:39 +24:37:11 &	18.0	& 1.27-1.00	& 160 &	HD 90183  \\
\hline
(31056)	& & & & & & & SA 93-101 \\
& 1996 RK25	& 10-28-2022	09:51 &	02:10:40 +00:16:44 &	17.4 &	1.11-1.07	& 68 &	HD 15166 	 \\
\hline 
(31499)	& & & & & & & SA 98-978 \\
& 1999 CS64	& 01-09-2021	12:25 &	07:14:09 +19:26:30 &	16.7 &	1.28-1.15	& 48 &	BD+161450 	 \\
\hline
(60378)	& & & & & & & SA 102-1081 \\
& 2000 AL165	& 01-23-2023	11:01 &	07:15:36 +26:56:30 &	17.7 &	1.32-1.08	& 72 &	HD 53532  \\
\hline
(52628)	& & & & & & & SA 105-56 \\
& 1997 WO3	& 05-15-2021	08:03 &	14:35:03 -19:06:01 &	17.9 &	1.39-1.29	& 52 &	HD 130958 	 \\
\hline
(52628)	& & & & & & & SA 105-56 \\
& 1997 WO3	& 05-19-2021	09:12 &	14:31:44 -18:46:50 &	18.0	& 1.58-1.28	& 44 &	HD 130958 	 \\
\hline 
(67244)	& 2000 EH58	& 08-13-2022	12:07 &	21:52:00 -12:58:43 &	18.3 &	1.77-1.08	& 52 &	HD 202497 \\
\hline
(76391)	& 2000 FP7	& 08-07-2022	09:26 &	21:14:31 -22:18:15 &	17.9 &	1.49-1.35	& 96 &	HD 202153  \\
\hline
(80052)	& & & & & & & SA 105-56 \\
& 1999 JV62	& 05-19-2021	11:30 &	15:02:19 -15:00:21 &	17.7 &	1.78-1.40	& 52 &	HD 135532 	 \\
\hline
(81819)	& 2000 KS35	& 08-20-2022	09:16 &	21:40:18 -16:34:04 &	17.4 &	1.29-1.24	& 52 & 	HD 206835 \\
\hline 
(85911)	& 1999 CY91	& 08-13-2022	09:47 &	21:09:41 -27:29:15 &	17.9 &	1.62-1.48	& 84 &	HD 224251  \\
\hline
\enddata
\tablecomments{For each asteroid, we report the name, provisional designation, observation start time, position of the asteroid in right ascension and declination at the start of observations, visible magnitude (as retrieved from JPL's Horizons online ephemeris service), the range of airmasses during the observation, total exposure time, and solar (if used) and local standard stars used. Observations with a single standard star were heavily affected by telluric absorptions when correction to the solar standard was performed, so for these observations we only used the local standard during flux to reflectance conversion (see text).}
\end{longdeluxetable}

\subsection{Lowell Discovery Telescope Observations}

We used the DeVeny Spectrograph's DV1 grating (R $\sim$450, \cite{DeVeny}) on the Lowell Discovery Telescope (LDT) to observe candidate asteroids in the visible (0.3 - 0.9 $\mu$m). Our observation strategy for the LDT is similar to the observing strategry for the IRTF described in the previous section. Observations of each target were taken at airmasses $<2.0$ with the slit aligned to the parallactic angle to reduce the chance of variation in the spectral slope due to differential atmospheric refraction. Each observation used a local standard star and a solar standard star to correct for the solar spectrum. Local standards were chosen by querying the SIMBAD astronomical database for bright (12 - 6 V mag) G dwarf stars within a 7 degree radius of the mean position of the asteroid on each night (\cite{simbad}). Solar standard stars were selected from a list of commonly used solar analogs (Bus, S. J., personal comm.). Observations of each asteroid were bracketed by observations of the local standard, returning to obtain local standard spectra after $\sim$ 60 minutes of integration time on each asteroid. Solar standards were observed once a night at low $< 1.5$ airmasses. This observing strategy allowed us to produce reflectance spectra for each object by averaging different calibrated versions of that object's spectrum from throughout each night. 

LDT data taken before July 2023 were extracted using SPECTROSCOPYPIPELINE (SP) developed for python by Maxime Devogele. SP performs bias subtraction, flat field correction, wavelength registration, telluric correction and conversion from absolute fluxes to relative reflectance (\cite{Devogele_SP}). SP also performs taxonomic classification by comparing asteroid spectra to the Bus-DeMeo types using a $\chi^2$ method. In July 2023, we updated our reduction process to use PypeIt (\cite{pypeit:joss_pub, pypeit:zenodo}) for bias subtraction, flat field correction, wavelength registration, and spectral extraction, with subsequent correction from absolute flux to relative reflectance performed in python. In both SP and PypeIt, this correction from absolute flux (uncorrected for solar color) to reflectance was accomplished using both the local and solar standards: first the asteroid spectrum was divided by the spectrum of the local standard to produce a locally corrected spectrum. Then, the solar analog spectrum was divided by the spectrum of the local standard to produce a solar comparison spectrum. The effect of this correction is to account for both variations in local atmospheric transparency (accomplished using the local standard) as well as correct for solar color (using the solar standard). Reflectance spectra taken using the LDT were normalized to 0.55 $\mu$m. The extracted spectra were smoothed and binned using a 10-point wide box filter prior to additional analysis. We observed a total of 21 targets identified in the SDSS MOC with the LDT. 

\begin{longdeluxetable}{llllcccl}
\tablecaption{Observation circumstances for asteroids observed with the Lowell Discovery Telescope \label{tab:LDT_obs}}
\tablewidth{0pt}
\tablehead{
 & \colhead{Provisional}	 & \colhead{Obs. Start} & 	& \colhead{Mag.}	& \colhead{Exp.}	& \colhead{Airmass}	& \colhead{Standard} \\
\colhead{Asteroid} & \colhead{Designation}	 & \colhead{(UTC)} & 	\colhead{RA/Dec}	& \colhead{(V)}	& \colhead{(min)}	& \colhead{Range}	& \colhead{Stars} 
}
\startdata
(203)  & & & & & & &	SA 105-56\\
Pompeja	& A879 SA &	03-04-2022 11:15 & 12:39:26 -05:29:17	& 13.2	& 2.5 &	1.53	& HD 110029 \\
\hline
(269) & & & & & & &	SA 110-361 \\
Justitia	& A887 SA &	05-24-2022 10:49 & 19:09:38 -13:41:10	& 12.8	& 6	& 1.51	& HD 180510	 \\
\hline 
(467) & & & & & & & SA 93-101 \\
Laura	& A901  AA	& 11-13-2021  06:55	& 06:24:42 +31:28:21	& 15.2	& 15	& 1.36-1.26 &	HD 259516 \\
\hline
(1947) & & & & & & & SA 115-271\\
Iso-Heikila & 1935 EA &	08-07-2021 	09:24 &	22:58:17 - 22:28:39	& 16.2	& 20	& 1.84	& HD 215393	\\
\hline 
(1947) & & & & & & &  SA 93-101\\
Iso-Heikila & 1935 EA &	12-02-2022 	04:02 &	02:58:13 +05:58:48	& 16.4	& 20	&1.25-1.21 &	HD 17762  \\
\hline
(3248) & & & & & & &  SA 105-56 \\
Farinella	& 1982 FK &	03-04-2022 	10:48 &	12:44:30 -02:28:41	& 16.4	& 15	& 1.38-1.35	& HD 111662	\\
\hline 
(7562) & & & & & & & SA 105-56 \\
Kagiroino-Oka &	1986 WO9 &	05-03-2021 	08:47 &	15:24:03 -15:36:23	& 17 &	30	& 1.68-1.60	& HD 131864 \\
\hline 
(8967) & & & & & & & SA 115-271\\
Calandra	& 4878 T-1	& 08-07-2021 	10:08 &	00:11:31 +02:04:35	& 17.6 &	30	& 1.19	& HD 1386 \\
\hline 
(13381)	& & & & & & & SA 93-101 \\
& 1998 WJ17	& 11-13-2021	02:01	& 00:03:10 +04:00:39	& 18.0 &	25	& 1.34-1.24	& HD 224251	\\
\hline 
(16551)	& & & & & & & SA 105-56 \\
& 1991 RT14	& 05-03-2021	04:25	& 10:24:24 +11:16:12	& 18.3 &	40	& 1.15-1.28	& HD89525	\\
\hline 
(17350)	& & & & & & & SA 110-361 \\
& 1968 OJ	& 05-24-2022	08:12	& 14:51:37 +04:23:53	& 18.4 &	48	& 1.54-1.33	& HD 128593	\\
\hline
(21867)	& & & & & & & SA 112-1333 \\
& 1999 TQ251 &	09-04-2022	03:33	& 19:56:23 -06:30:25	& 18.3 &	58	& 1.37-1.32	& HD 187490	\\
\hline
(22689)	& & & & & & & SA 93-101 \\
& 1998 QQ84	& 11-13-2021	08:03	& 07:06:40  +13:06:18	& 18.3 &	80	& 1.43-1.15	& HD 52634	\\
\hline 
(23248) & & & & & & & SA 93-101 \\
Batchelor	& 2000 WW178	& 11-13-2021	09:58	& 09:08:29 +22:08:22	& 19.0 &	90	& 1.35-1.08 &	HD 76752 \\
\hline
(26895) & & & & & & & SA 107-684 \\
& 1995 MC & 07-09-2023 09:01 & 22:12:07 -03:58:10 & 16.2 & 15 & 1.39-1.35 & HD 210335\\
\hline 
(27378)	& & & & & & &  SA 105-56\\
& 2000 EG55	& 03-04-2022	09:50	& 10:26:07 +25:10:37	& 18.0 &	40	& 1.42-1.25	& HD 90183	\\
\hline 
(28368) & & & & & & & SA 110-361 \\
& 1999 GW18 & 07-09-2023 03:56 & 15:52:47 -13:56:26 & 18.5 & 100 & 1.52 - 2.04 & HD 142801 \\
\hline
(31056)	& & & & & & & SA 93-101 \\
& 1996 RK25	& 10-28-2022	08:14	& 01:55:12 -02:40:58	& 17.4 &	30	& 1.38-1.32 &	HD 11752 \\
\hline 
(31786)	& & & & & & & Hya 64\\
& 1999 KO13	& 10-28-2022	09:03	& 03:01:53 +28:37:45	& 17.5 &	25	& 1.05-1.03 &	HD 19823 \\
\hline
 \\
(33693)	& & & & & & &  SA 93-101 \\
& 1999 KA	& 11-13-2021	06:07	& 03:29:42 -01:59:53	& 17.7 &	35	& 1.33-1.27 &	HD 21316 \\
\hline 
(36249)	& & & & & & & SA 110-361 \\
& 1999 VT178	& 05-24-2022	09:27	& 17:41:11 -06:01:47	& 18.4 &	60	& 1.45-1.33 &	HD 159006 \\
\hline
(40131)	& & & & & & &  SA 105-56 \\
& 1998 QJ48	& 03-04-2022	09:01	& 11:16:04 -01:01:31	& 17.6 &	30	& 1.34-1.29	& HD 97275  \\
\hline 
(67141)	& & & & & & & SA 115-271 \\
& 2000 AC169 &	08-07-2021	08:41	& 21:01:32 - 12:59:20	& 17.7 &	30	& 1.67-1.57	& HD 197759	\\
\hline 
(81819)	& & & & & & &  SA 112-1333 \\
& 2000 KS35 &	09-04-2022	06:26	& 21:41:30 -18:25:11	& 17.9 &	44	& 1.73-1.66	& HD 205291	\\
\hline 
(85911)	& & & & & & & SA 112-133 \\
& 1999 CY91	& 09-04-2022	05:04	& 21:35:30 -15:26:17	& 18.4 &	60	& 1.64-1.56	& HD 203812  \\
\hline
(106063)	& & & & & & &  Hya 64\\
& 2000 SR319 &	10-28-2022	09:43	& 04:00:56 +37:06:47	& 17.2&	25	& 1.03-1.01	& HD 26182 \\
\hline 
(158762)	& & & & & & &  SA 93-101 \\
& 2003 RS	& 11-13-2021	05:06	& 03:07:20 +25:46:45	& 18.0 &	35	& 1.12-1.05	& HD 19445	\\
\hline 
(246945)	& & & & & & &  SA 93-101 \\
& 1999 RP116 &	11-13-2021	02:57	& 00:51:20 +38.:38:41	& 18.5 &	70	& 1.09-1.01	& HD 6664 \\
\enddata
\tablecomments{For each asteroid, we report the name, provisional designation, observation start time, position of the asteroid in right ascension and declination at the start of observations, visible magnitude (as retrieved by JPL's Horizons online ephemeris service), the range of airmasses during the observation, total exposure time, and solar and local standard stars used.}
\end{longdeluxetable}

\subsection{Analysis and Spectral Classification}

Following spectral extraction and conversion from flux to relative reflectance, we evaluated each spectrum using various classification methods. First, we determined the spectral slopes of each asteroid by performing a linear fit on each normalized spectrum. To estimate errors in spectral slope for each spectrum, we used a Monte Carlo method, generating 1000 synthetic spectra per measurement by drawing reflectance values from a Gaussian distribution centered at the mean reflectance value at each wavelength and error equal to the associated error in reflectance. For each synthetic spectrum, we computed the spectral slope of the best fitting line, taking the mean spectral slope of all 1000 synthetic spectra as the spectral slope measurement, and the standard deviation of the slope measurements as the error. To improve the accuracy of these linear fits, we excluded regions of high noise from the linear fitting process. For the LDT data, we excluded wavelengths $< 0.35 \mu$m and $ > 1.0 \mu$m, fitting slope over the $0.35 - 1.0 \mu$m range. For the IRTF data, we excluded wavelengths $> 2.4 \mu$m, fitting slope over the $0.8-2.4 \mu$m range. The resulting linear fit was assessed visually against each spectrum to ensure an appropriate fit. Given a linear fit, we calculate spectral slope as the predicted increase in spectral slope over the 0.55 - 0.65 $\mu$m range relative to reflectance at 0.55 $\mu$m. That is, spectral slope $S$ as a percentage is given by the equation

\begin{equation}
    S(\%) = \frac{R(0.65 \mu m) - R(0.55 \mu m)}{R(0.55 \mu m)} \times 100.
\end{equation}

\noindent $R(x)$ is the predicted solar corrected reflectance at wavelength $x$. For IRTF observations, this calculation necessarily requires extrapolating the average spectral slope in the $0.8 - 2.4 \mu$m range to the visible to compute predicted reflectances at 0.65 and 0.55 $\mu$m. We also calculated the slope of the R group average spectra from \cite{Emery_2010} over the $0.35 - 1.0$ $\mu$m and $0.8 - 2.4$ $\mu$m ranges. Note that, because many asteroid spectra deviate from perfectly linear, the spectral slopes measured over the visible range ($0.35 - 1.0$ $\mu$m) and the near-infrared range ($0.8 - 2.4$ $\mu$ m) are not necessarily equal. In particular, the typical D-type asteroid in the Bus-DeMeo taxonomy may show a shallowing of infrared slopes at wavelengths longer than ~1.5 $\mu$m (\cite{demeo2009}). For the average R group Trojan, we find the average slope from $0.35 - 1.0$ $\mu$m, $S_{0.35 - 1.0 \mu m} = 10.7 \%$ and the average slope from $0.8 - 2.4$ $\mu$m, $S_{0.8 - 2.4 \mu m} =4.99 \% $.

We also classified the asteroids using the Bus-DeMeo taxonomy (\cite{demeo2009, bus1999, bus2002}). For the IRTF spectra, we used the Bus-DeMeo Classification Web tool developed by Stephen M. Slivan to determine taxonomic type. This method uses principle component analysis (PCA) to classify asteroid spectra and provides several possible ``best fit'' classifications. For spectra classified using the Bus-DeMeo classifier, we report the best fit classification as measured by the lowest absolute residual. The Classification Web tool requires either full VNIR coverage (0.45 to 2.45 $\mu$m) or NIR coverage (0.85 to 2.45 $\mu$m) to classify an asteroid spectrum. Therefore, for objects that were only observed with the LDT, we used the taxonomic classifications provided by comparing the normalized asteroid spectrum to the normalized spectra of the collection of channel averages for each spectral type presented in \cite{demeo2009} and identifying the best fit as the spectral class with the smallest $\chi^2$-goodness-of-fit metric. We note that spectral classifications based only on visible data are less reliable than those taken in the near infrared (or across both the visible and near infrared) as many characteristic spectral features, such the subtle broad but shallow 1.0 - 1.3 $\mu$m feature that distinguishes C-type asteroids from X-type asteroids (\cite{demeo2009}), are either absent from or located in low signal-to-noise regions of the LDT spectra. 

\begin{figure}
    \includegraphics[width=\linewidth]{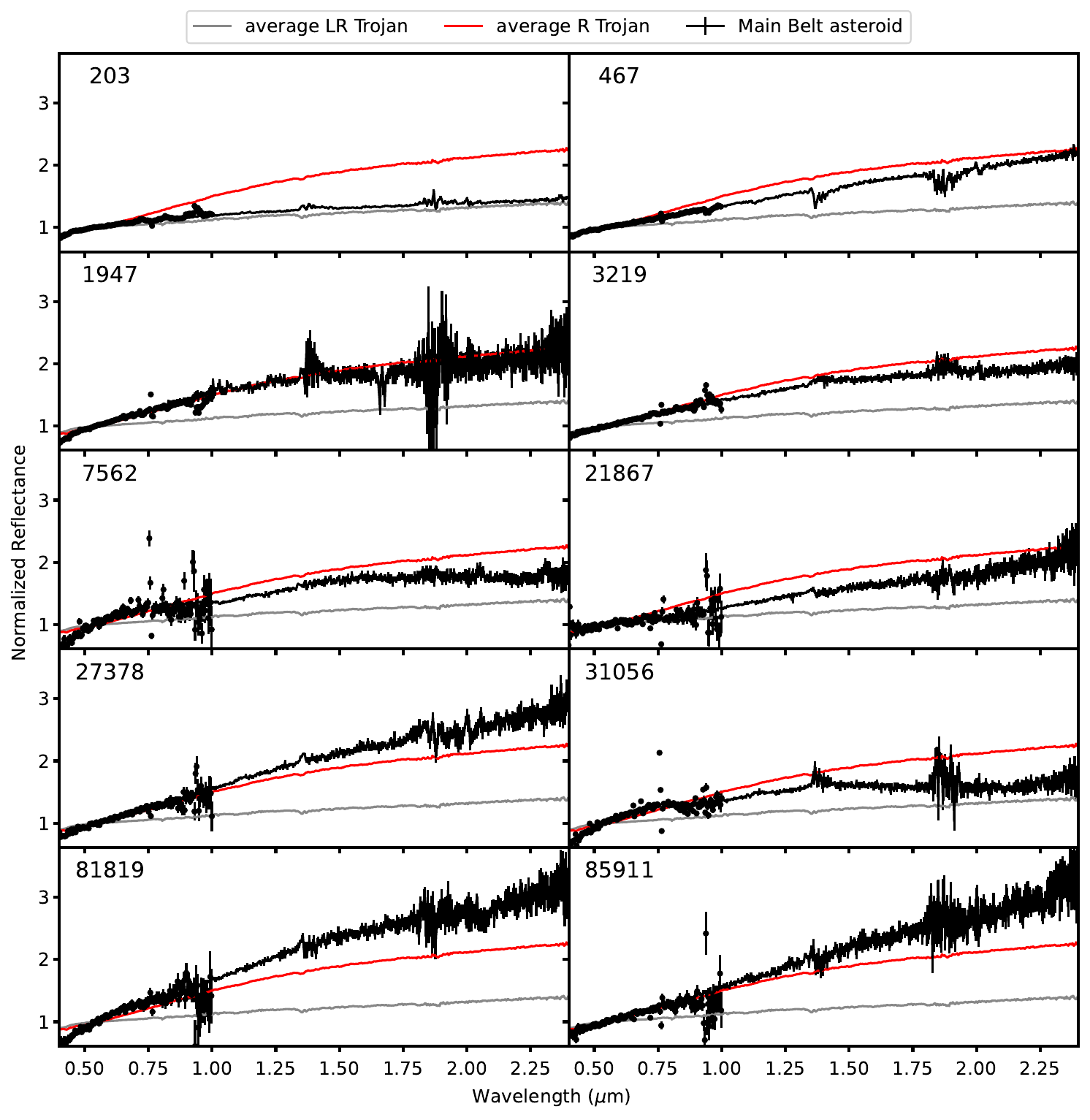}
    \caption{Visible and near-infrared spectra of all asteroids observed with both the LDT and IRTF in this survey. All spectra are normalized to 1 at 0.55 $\mu$m and displayed at the same scale to enable direct visual comparison. Each spectrum is plotted alongside the average R and LR group Trojan reflectance spectrum from \cite{Emery_2010}. \label{fig:full_all}}
\end{figure}

Ten of our targets have full wavelength coverage from both the IRTF and the LDT. For these targets, we used the linear fits we derived to predict reflectance at 0.8 $\mu$m, in the region of overlap between the LDT and IRTF data, then scaled the IRTF spectra to the predicted visible flux. The effect of this scaling is to normalize the entire spectrum (0.3 - 2.5 $\mu$m) to a value of 1 at 0.55 $\mu$m. When combining the data, we excluded LDT data longwards of 1.0 $\mu$m due to the greater noise in this region, which was well-characterized in the IRTF spectra. These full wavelength spectra (see Figure \ref{fig:full_all}) can be classified using the Bus-DeMeo Classification Web tool. We prioritize reporting classifications derived using the full wavelength coverage when these data were available. For all classifications that relied on the Bus-DeMeo Classification Web tool, we did not assign a class to asteroids with absolute residuals $>0.1$ as none of the Bus-DeMeo taxonomic classes provided a good match to the spectrum. We note that this combination method does not take into account phase reddening or other time-variable factors, such as rotational phase, that may affect an asteroid's spectral appearance.

For spectra obtained with the IRTF, we also calculated synthetic color indices in the near-infrared. In \cite{Emery_2010}, the LR and R group Trojans were found to form two distinct clusters in (0.85 $\mu$m - J) and (J - K) color space. We computed synthetic color indices by averaging reflectance values within $\pm 0.1 \mu$m of each band center, e.g. 0.85 $\mu$m, 1.25 $\mu$m (J band), and 2.2 $\mu$m (K band). Average reflectances $R(0.85 \mu m), R(1.25 \mu m) \approx J,$ and $ R(2.2 \mu m) \approx K$ were then converted to differences in magnitudes to obtain synthetic color index estimations using 

\begin{equation} \label{85_J} 
   (0.85 - J) = -2.5 \log_{10} \left( \frac{R(0.85 \mu m)}{R(1.25 \mu m)} \right) 
\end{equation}

\noindent and 

\begin{equation} \label{J_K} 
   (J - K) = -2.5 \log_{10} \left( \frac{R(1.25 \mu m)}{R(2.2 \mu m) } \right).
\end{equation}

\noindent Here again, $R(x)$ represents the normalized reflectance measured at wavelength $x$. We note that since these color indices are calculated from a reflectance spectrum (e.g., a spectrum from which the solar spectrum has already been divided), the color indices derived are already corrected for solar colors. We used the uncertainty in spectral slope value of $\sim 4 \%/\mu$m reported in \cite{marsset2020} for absolute uncertainty in spectral slope expected for IRTF measurements to estimate the errors for each color index. The expected error in color index due to slope uncertainty was added in quadrature to the propagated errors from the averaging steps described above. For most asteroids, the error due to absolute uncertainty in spectral slopes is the dominant contribution to the error in computed color indices.

\section{Results}

\begin{deluxetable}{llccccc}

\tablecaption{Slope measurements and classifications for asteroids observed with the IRTF and LDT \label{tab:all_res}}
\tablewidth{0pt}
\tablehead{
 & Provisional & $S_{0.35-0.7 \mu m}$ &$S_{0.8-2.4 \mu m} $ & & & Bus-DeMeo \\
 Asteroid & Designation & ($\%/0.1 \mu m$) & ($\%/0.1 \mu m$) & 0.85 - J & J - K & classification
}
\startdata
(203) Pompeja	& A879 SA &	6.50 $\pm$ 0.07 &	1.85 $\pm$ 0.01 &	0.10 $\pm$ 0.02 & 	0.13 $\pm$ 0.04 &	X \\
(269) Justitia	& A887 SA &	\textit{18.7 $\pm$ 0.06} &  &  &  & D\textdagger \\
(467) Laura	    & A901 AA & 8.19 $\pm$ 0.06	 & \textit{ 5.92 $\pm$ 0.01} &	0.24 $\pm$ 0.02	& 0.34 $\pm$ 0.04 &	D \\
(1947) Iso-Heikila & 1935 EA &	\textit{12.6 $\pm$ 0.28} & 4.17 $\pm$ 0.09 & 0.25 $\pm$ 0.02 & 0.21 $\pm$ 0.04 & D \\
(1947) Iso-Heikila & 1935 EA &	\textit{11.8 $\pm$ 0.17} &		&  & 		& D\textdagger \\
(3248) Farinella & 1982 FK	& 10.4 $\pm$ 0.13 &	3.91 $\pm$ 0.03 & 0.24 $\pm$ 0.02 &	0.20 $\pm$ 0.04 &	D\\
(5819) Lauretta	& 1989 UZ4	&  &	1.97 $\pm$ 0.02	& 0.18 $\pm$ 0.02 & 0.08 $\pm$ 0.04 & S* \\
(7562) Kagiroino-Oka &	1986 WO9 &	\textit{13.4 $\pm$ 0.82} & 2.78 $\pm$ 0.03 & 0.16 $\pm$ 0.02 & 0.14 $\pm$ 0.04	& - \\
(8967) Calandra	& 4878 T-1 & 6.64 $\pm$ 0.73 &  &  &  & K\textdagger \\
(9934) Caccioppoli & 1985 UC & 		& \textit{5.81 $\pm$ 0.03} & 0.24 $\pm$ 0.02 &	0.32 $\pm$ 0.04 & D*  \\
(13381) &	1998 WJ17 &  9.08 $\pm$ 1.9 &	 &  &  & D\textdagger \\
(16551)	& 1991 RT14 &	10.0 $\pm$ 0.65 &	 &	 &	 &	A\textdagger \\
(17350)	& 1968 OJ &	7.15 $\pm$ 0.83	& 	& 	&  &  S\textdagger \\
(21867)	& 1999 TQ251 &	5.44 $\pm$ 0.70 &	\textit{6.18 $\pm$ 0.07} & 0.26 $\pm$ 0.02 & 0.33 $\pm$ 0.04 &	- \\
(22110) & 2000 QR7 & 		& \textit{6.96 $\pm$ 0.14} & 0.23 $\pm$ 0.02 &	0.38 $\pm$ 0.04 &-* \\
(22422) Kenmount Hill &	1995 YO5 &		& \textit{9.91 $\pm$ 0.14} & 0.35 $\pm$ 0.02 &	0.45 $\pm$ 0.04 & -* \\
(22689) &	1998 QQ84 &	-1.31 $\pm$ 0.49 &  &  &  & Ch\textdagger \\
(23248) Batchelor	& 2000 WW178 &	\textit{20.6 $\pm$ 0.42}	&  &  & 	& A\textdagger \\
(23690) &	1997 JD14 &		& \textit{6.63 $\pm$ 0.24} & 0.31 $\pm$ 0.02 &	0.38 $\pm$ 0.04 &	D* \\
(25835) Tomzega & 2000 EO20 &	 & \textit{7.44 $\pm$ 0.22} &	0.32 $\pm$ 0.03 & 0.34 $\pm$ 0.04 &	-* \\
(26895)	& 1995 MC &	\textit{14.8 $\pm$ 0.06} &	 &	 &	 &	T\textdagger\\
(27378) & 2000 EG55 & \textit{11.9 $\pm$ 0.52} & \textit{8.92 $\pm$ 0.07} &	0.36 $\pm$ 0.02 & 0.39 $\pm$ 0.04 & -  \\
(28368) & 1999 GW18	& 10.7 $\pm$ 1.73 &	 &  &	 &	D\textdagger \\
(31056) & 1996 RK25 & \textit{12.3 $\pm$ 0.22} & 1.71 $\pm$ 0.04 & 0.16 $\pm$ 0.02 & 0.05 $\pm$ 0.04 & S \\
(31499) & 1999 CS64 &		& \textit{6.60 $\pm$ 0.06} & 0.33 $\pm$ 0.02 &	0.31 $\pm$ 0.04 & D* \\
(31786)	& 1999 KO13	& 10.5 $\pm$ 0.20 &	 &	 &  &	D\textdagger \\
(33693)	& 1999 KA &	8.75 $\pm$ 0.42 & &  &  &	D\textdagger \\
(36249)	& 1999 VT178 &	\textit{11.3 $\pm$ 0.55} &	 &	 &	 &	L\textdagger \\
(40131) & 1998 QJ48	& 7.93 $\pm$ 0.31 &	 & &	 &	D\textdagger \\
(52628) & 1997 WO3 &  & 0.34 $\pm$ 0.07 &	0.03 $\pm$ 0.02 &	0.02 $\pm$ 0.04 &	L* \\
(60378)	& 2000 AL165 &		& 2.00 $\pm$ 0.04 &	0.16 $\pm$ 0.02 &	0.08 $\pm$ 0.04 &	S* \\
(67141) & 2000 AC169 &	\textit{10.9 $\pm$ 0.29} &	 &	 &  & L\textdagger \\
(67244) & 2000 EH58 &		& \textit{6.97 $\pm$ 0.12} & 0.32 $\pm$ 0.02 &	0.31 $\pm$ 0.04 & -* \\
(76391) & 2000 FP7 &		& \textit{7.57 $\pm$ 0.07} & 0.34 $\pm$ 0.02 &	0.34 $\pm$ 0.04 & -* \\
(80052)	& 1999 JV62 &	 	& 1.16 $\pm$ 0.09 &	0.09 $\pm$ 0.02 &	0.07 $\pm$ 0.04 &	S* \\
(81819) & 2000 KS35 &	\textit{16.0 $\pm$ 1.04 } &	\textit{9.36 $\pm$ 0.10} &	0.39 $\pm$ 0.02 &	0.40 $\pm$ 0.04 &	-\\
(85911) & 1999 CY91 &	\textit{11.2 $\pm$ 0.77} & \textit{12.1 $\pm$ 0.15} & 0.38 $\pm$ 0.02 &	0.48 $\pm$ 0.04 & - \\
(106063) &	2000 SR319 &	\textit{11.3 $\pm$ 0.15} &	 &	 &	 &	L\textdagger \\
(158762) &	2003 RS &	7.85 $\pm$ 1.74 &	 &	 &		& D\textdagger \\
(246945) &	1999 RP116 &	7.78 $\pm$ 2.89 &		& & 	&	D\textdagger \\
\enddata
\tablecomments{For each asteroid, we report (where applicable) the name, provisional designation, spectral slope in the visible, spectral slope in the infrared, 0.85 - J color, K- J color, and Bus-DeMeo classification. \textit{Italicized} entries indicate asteroids with a steep spectral slope measurement, e.g. those whose spectral slopes exceed the average spectral slope of an R group Trojan in the visible (10.7 \%/0.1 $\mu$m) or in the infrared (4.99 \%/0.1 $\mu$m). Note that the Bus-DeMeo classification in some cases is ambiguous; for these asteroids, spectral classification was disambiguated by choosing the classification with the lowest absolute residual. Asteroids marked with a class of `-' were not assigned a classification within the Bus-DeMeo system as none of the taxonomic classes in that system gave residuals $<0.1$. Classifications marked with an asterisk (*) are based on IR-only data, while those marked with a dagger (\textdagger) are based on visible-only data and should be taken as preliminary--see Section 2.4 for details on classification.}
\end{deluxetable}

\begin{figure}
\includegraphics[width=\linewidth]{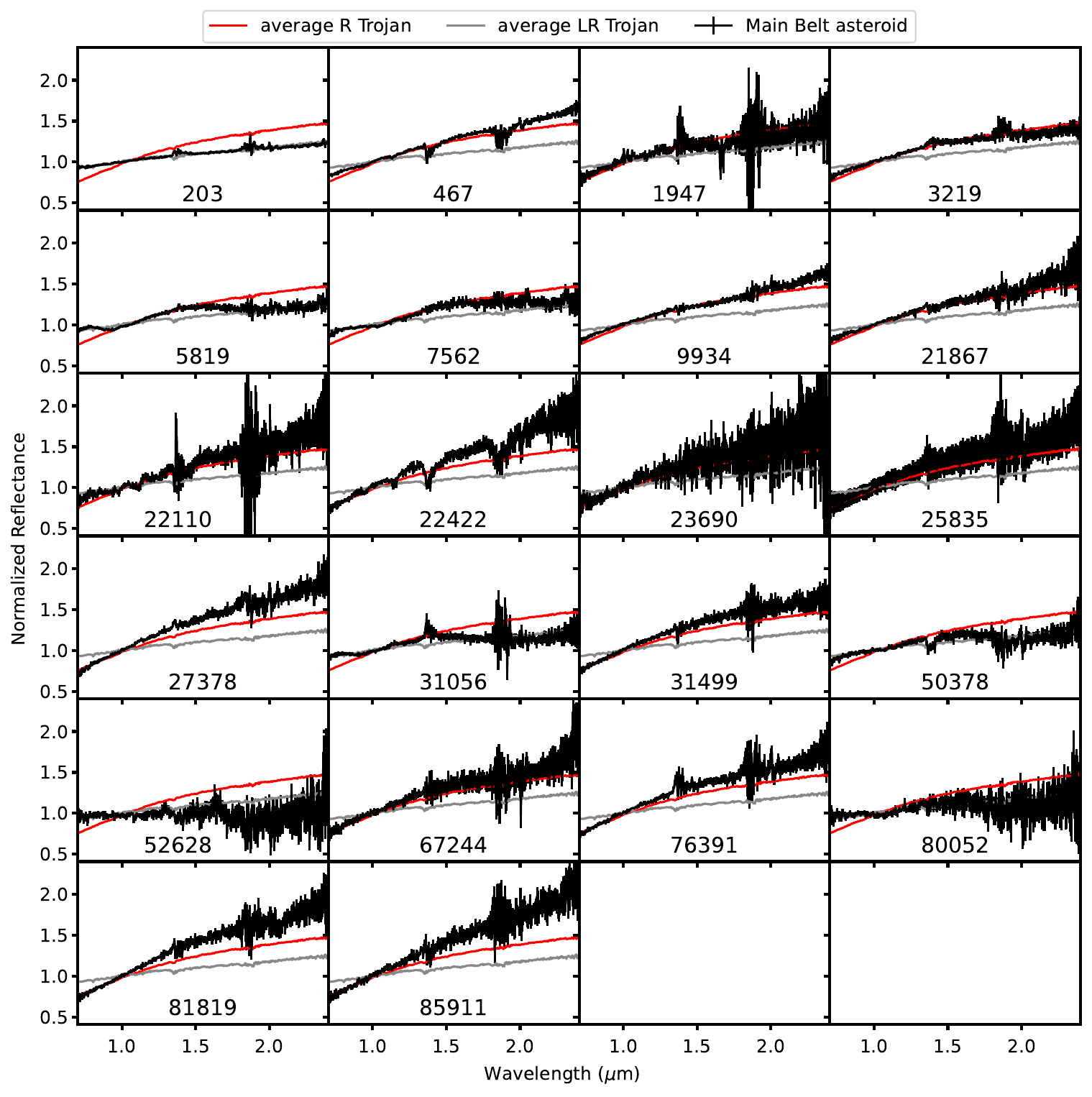}
\caption{Near-infrared reflectance spectra of all objects observed with the IRTF. All spectra are normalized to 1 at 1 $\mu$m and displayed at the same scale to enable direct visual comparison. Each spectrum is plotted alongside the average R and LR group Trojan reflectance spectrum from \cite{Emery_2010}. \label{fig:IRTF_all}}
\end{figure}

We report the average spectral slopes over 0.8 - 2.4 $\mu$m, Bus-DeMeo classification, 0.85 - J and J - K color indices for all asteroids observed using the NASA IRTF in Table \ref{tab:all_res}. The individual spectrum for each asteroid is also displayed alongside the average spectra for the R and LR group Trojans from \cite{Emery_2010} in Figure \ref{fig:IRTF_all}. We observed a total of 22 asteroids with the IRTF, 17 of which were drawn from the final, most restrictive list of SDSS MOC `red' candidate objects with multiple steep slope measurements and/or low albedos. Four of the other asteroids were drawn from the backup list of SDSS MOC candidates that had at least one steep slope observation, but did not consistently show high slopes, and the fifth was (203) Pompeja, which was observed after its steep spectral slope was reported in \cite{hasegawa2021}. Of the 17 SDSS candidates, 11 asteroids have slopes exceeding the average R group Trojan and 5 are classified as D types.

\begin{figure}
\centering
\includegraphics[width=0.75\linewidth]{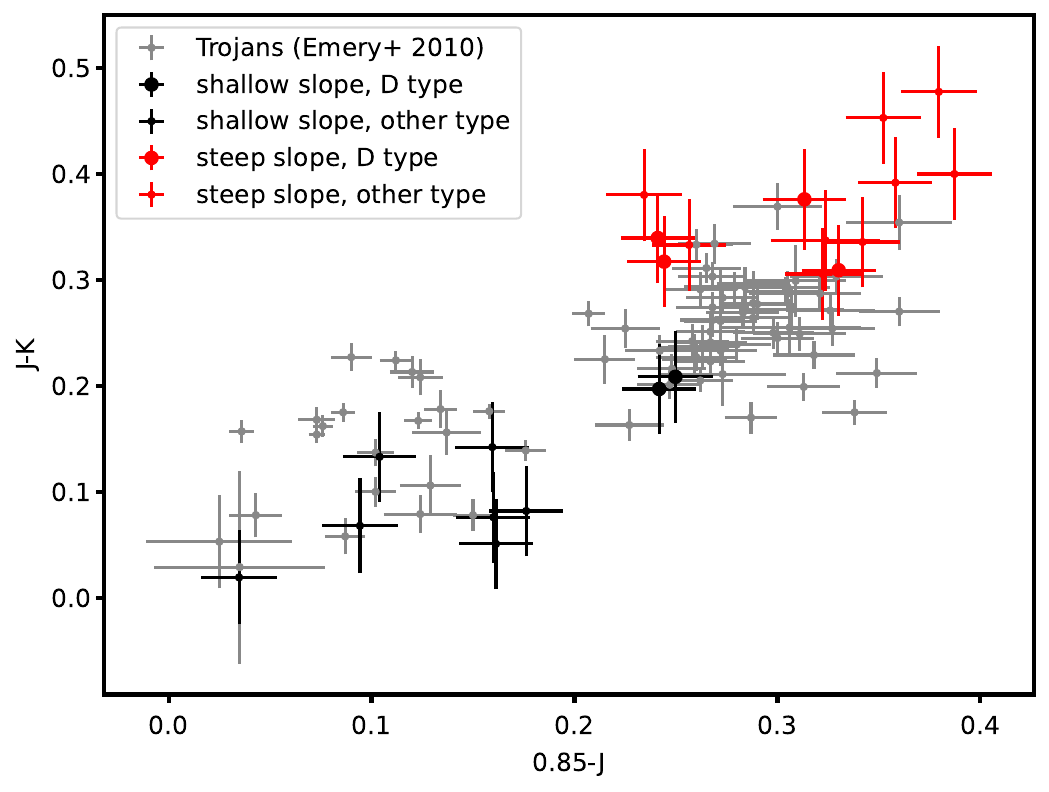}
\caption{0.85 - J and J - K color indices of all asteroids observed with the NASA IRTF. Asteroids with spectral slopes steeper than the average R group Trojan are plotted in red, those with shallower slopes are plotted in black. Asteroids classified as D types are plotted with larger markers than those classified as all other types. The color indices of LR (lower left cluster, grey) and R (upper left cluster, grey) groups from \cite{Emery_2010} are plotted for comparison. We draw attention to the asteroids that have higher J - K color values than would be predicted based on their 0.85 - J color if this population was Trojan-like that plot above the main cluster of R group Trojans. \label{fig:color_index}}
\end{figure}

Examining the 0.85 - J and J - K color indices of the asteroids observed with the IRTF in comparison to the Jupiter Trojans from \cite{Emery_2010} (See Figure \ref{fig:color_index}), we note that many of the D-types we identified plot within the main cluster of R group Trojans with 0.85 - J color indices between 0.2 and 0.35 and J - K color indices between 0.15 and 0.4. We note three steeply sloped asteroids with higher J - K color indices than would be predicted from their 0.85 - J values, that is, they plot above the cluster of R group Trojans in Figure \ref{fig:color_index and are redder at longer wavelengths than the majority of R group Trojans. These asteroids are (85911) 1999 CY91, (22422) 1995 YO5, and (22110) 2000 QR7.} Spectrally, they lack the characteristic ``rollover'' from steeper red slopes at short wavelengths to shallower red slopes at longer wavelengths (beyond $\sim$ 1.5 $\mu$m), showing approximately linearly increasing slopes in the near-infrared, which is reflected by their location within the 0.85 - J and J - K color index space. 

\begin{figure}
\includegraphics[width=\linewidth]{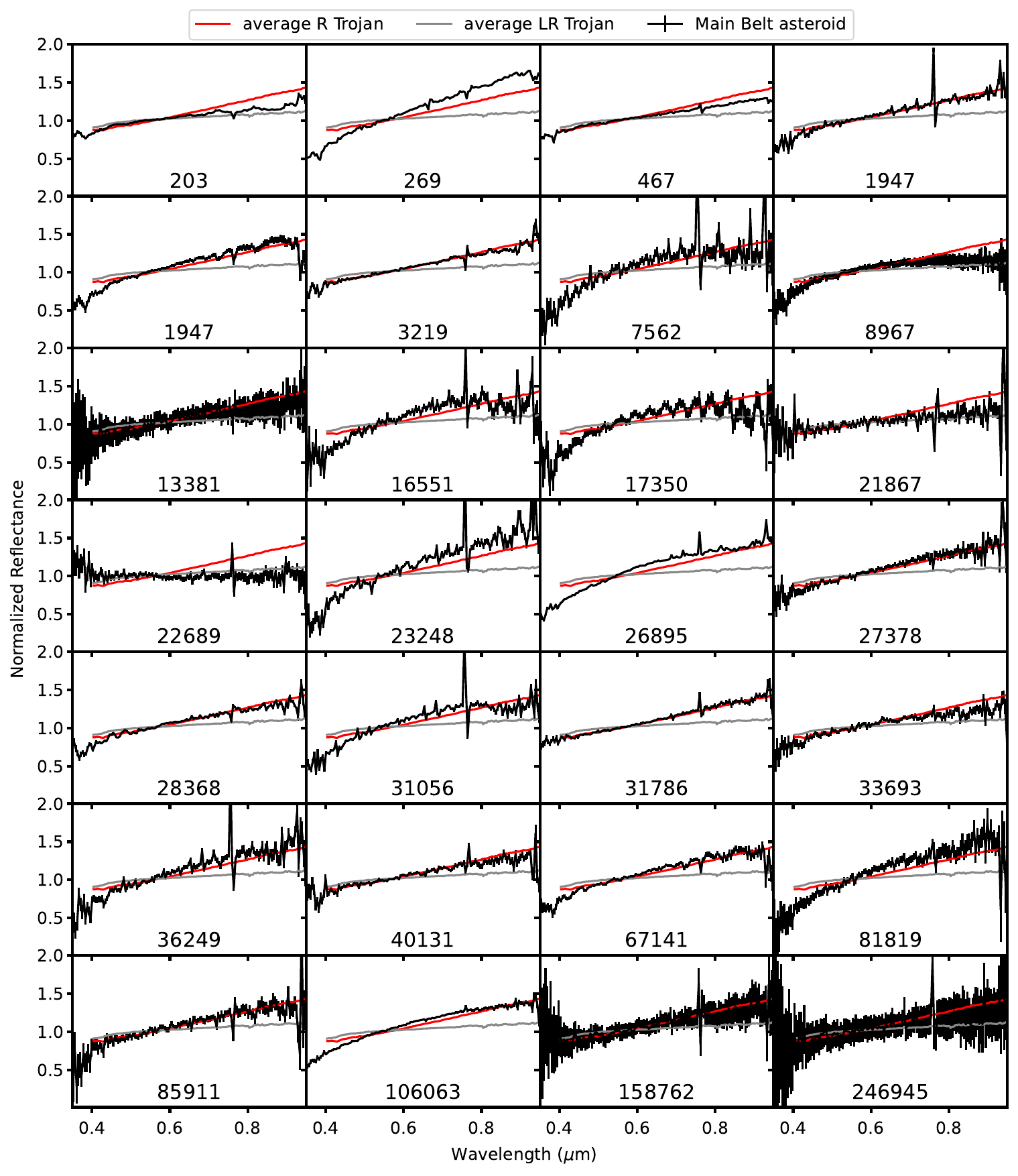}
\caption{Visible reflectance spectra of all objects observed with the LDT. Spectra from the 2021 and 2022 apparitions of (1947) Iso-Heikila are plotted seperately. All spectra are normalized to 1 at 0.55 $\mu$m and displayed at the same scale to enable direct visual comparison. Each spectrum is plotted alongside the average R and LR group Trojan reflectance spectrum from \cite{Emery_2010}. \label{fig:LDT_all}}
\end{figure}

We report the average spectral slopes over 0.35 - 1.0 $\mu$m and Bus-DeMeo classification for all asteroids observed using the LDT in Table \ref{tab:all_res}. The individual spectrum for each asteroid is also displayed alongside the average spectra for the R and LR group Trojans from \cite{Emery_2010} in Figure \ref{fig:LDT_all}. We observed a total of 27 asteroids with the LDT, 21 of which were drawn from the final, most restrictive list of SDSS MOC `red' candidate objects with multiple steep slope measurements and/or low albedos. Four of the other asteroids were drawn from the backup list of SDSS MOC candidates that had at least one steep slope observation, but did not consistently show steep slopes, and the remaining two were (203) Pompeja and (269) Justitia, which were observed after their steep spectral slopes were reported in \cite{hasegawa2021}. Of the 21 SDSS candidates, 11 asteroids have slopes exceeding the average R group Trojan and 9 are classified as D types. 

\section{Discussion}

The results of this survey can be extrapolated to the rest of the SDSS MOC catalog to estimate the prevalence of various populations of red asteroids in the Main Belt. We estimate the prevalence of a population of red asteroids by calculating the proportion of SDSS MOC-identified candidate red objects we confirmed to be a part of that population, then calculate how many asteroids in the entire SDSS MOC we expect would be that population, assuming the asteroids we observed are a representative sample. For the purposes of this extrapolation, we examine two different sub-populations of `red' asteroids: 

\begin{itemize}
    \item \textbf{D-type asteroids.} This category includes all asteroids classified as D-type. For asteroids with multiple observations, we prioritize the taxonomic type determined using the full-wavelength range (0.35 - 2.5 $\mu$m), otherwise, classifications are based on all available spectral data. 
    \item \textbf{Steep sloped asteroids.} This category includes all asteroids with measured slopes in the visible and/or near-infrared that exceed the spectral slope of the average R group Trojan over the observed wavelength range. These asteroids may be of any taxonomic type. 
\end{itemize}

\begin{deluxetable}{llccc}
\tablecaption{Confirmation Rates of 'Red' Asteroid Subpopulations \label{tab:sub-pop}}
\tablewidth{0pt}
\tablehead{
 & & Number & Confirmation & Extrapolated \\
 Instrument & Sub-population & Observed & Rate ($\%$) & Population \\
}
\startdata
 IRTF & D-type	& 5 & 29 & 18 \\
 & Steep Sloped & 11 & 65 & 39 \\
 &	Total SDSS Candidates & 17 \\	
\hline
 LDT & D-type & 9 & 43 & 26 \\
 &	Steep Sloped & 11 & 52 & 31 \\
 &	Total SDSS Candidates & 21 \\
\hline
IRTF or LDT &	D-type & 	12	& 38  & 23  \\
 & Steep Sloped & 19 &  59 &  36 \\
 & Total SDSS Candidates &	32	
\enddata
\tablecomments{We report the number of confirmed SDSS candidate objects in each `red' sub-population (see text) among the IRTF, LDT, and complete spectral datasets. Confirmation rates are calculated by dividing the number of observed SDSS MOC `red' candidate objects belonging to each subpopulatation by the total number of observed SDSS MOC `red' candidate objects observed using each facility. Extrapolated population size is given by multiplying these confirmation rates by the total number (60) of SDSS MOC objects that met our strictest criteria for inclusion as `red' candidate objects.}
\end{deluxetable}

We summarize the confirmation rates for each of these asteroid sub-populations in Table \ref{tab:sub-pop}. Our confirmation rate for D-type asteroids based on our selection criteria is $\sim 40 \%$. This confirmation rate is significantly higher than the $\sim 20\%$ confirmation rate of \cite{demeo2014_dtype} for candidate D-type inner Main Belt asteroids identified using SDSS MOC spectrophotometry. This may be due to increased prevelance of D-type Main Belt asteroids with semimajor axes  $>2.5 $ AU, but we note that our criteria for inclusion are stricter than those used in \cite{demeo2014_dtype}, as we require our candidate asteroids to have spectral slopes exceeding that of the average R group Trojan asteroid and low geometric albedos (when available). In \cite{demeo2014_dtype}, the criteria for inclusion was at least one D-type classification. As the average R group Trojan has a steeper spectral slope than the typical D-type asteroid (\cite{Emery_2010}) and we required an object to have multiple steeply sloped observations and/or a low albedo to be included in our list of candidates, it is more likely that the more stringent inclusion criteria used in this paper led to the higher confirmation rate for D-type asteroids. Our confirmation rate for steep sloped asteroids is higher ($\sim 50\%$), reflecting the fact that not all steep sloped asteroids classified as D-types. 

Not all of the asteroids observed in this study were identified as candidates according to the strictest set of criteria. We observed five asteroids (four among the LDT observations, and four among the IRTF observations) that were drawn from a list of backup objects with at least one steep slope observation but displayed high variability in slope in the MOC and/or high ($>0.15$) albedos. Due to the small sample size of objects from this list, we refrain from drawing any conclusions about the confirmation rates of D-type, steep slope, or Main Belt Trojan analog asteroids among asteroids with inconsistently red slopes and/or high albedos. However, we note that three of the five asteroids from the backup list had steep slopes exceeding the spectral slope of the average R group Trojan, and one, (467) Laura, also classified as a D-type.

\begin{figure}
\centering
\includegraphics[width=0.75\linewidth]{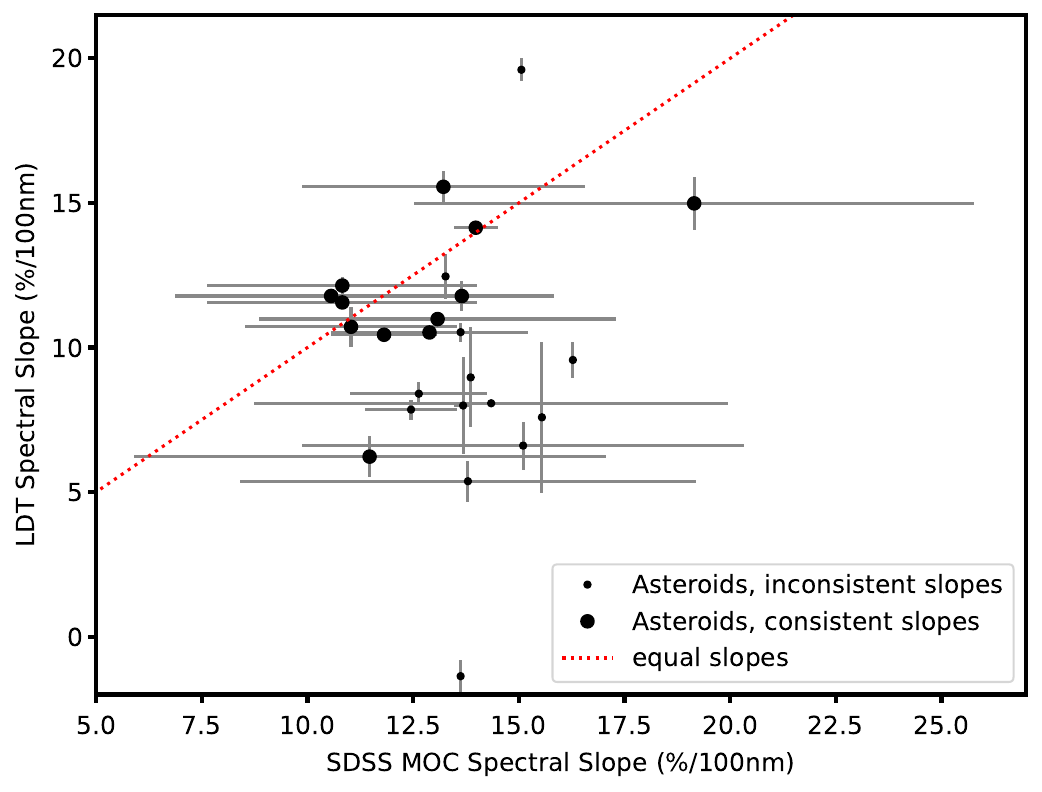}
\caption{Comparison of asteroid spectral slopes measured using the SDSS MOC (x-axis) and Lowell Discovery Telescope (y-axis). Error bars in the x-direction represent the full range of slopes measured in the SDSS MOC for asteroids with multiple slope measurements. Error bars in the y-direction represent errors in slope corresponding to measurement uncertainty associated with the Lowell Discovery Telescope spectra. The dotted line represents the location of the 1:1 line where SDSS MOC and LDT slope measurements would be equal. Those asteroids plotted with a large marker show an overlap in measured slope ranges in the SDSS MOC and LDT data, those asteroids plotted with a small marker do now show an overlap in measured slope ranges between the two datasets. \label{fig:slope_compare}}
\end{figure}

Our confirmation rates can be used to estimate the reliability of using SDSS MOC data to measure the spectral slopes of asteroids (See Figure \ref{fig:slope_compare}). We observed a total of 25 asteroids with colors available in the SDSS MOC with the LDT. Comparing the range of slopes measured in the MOC to the range of slopes associated with measurement error from the LDT, we find that 10 asteroids show an overlap in measured slope ranges. A total of 12 asteroids have shallower (e.g. bluer) slopes in the LDT data than predicted by the MOC and 1 asteroid has a steeper (e.g. redder) slope in the LDT data than predicted by the MOC. This distribution suggests that SDSS MOC data tends to over-estimate spectral slopes. Discrepancies in slope between the LDT and SDSS MOC may be due to rotational variability in slope intrinsic to the object, like the variability in slope observed by \cite{SouzaFeliciano_2020} among the Jupiter Trojans. However, the observed bias of the SDSS MOC observations towards overestimating the red slopes of asteroids suggests a bias in experimental design. To identify steeply red sloped candidates, we corrected SDSS MOC photometry to solar colors, which does not account for the potential reddening due to atmospheric extinction. Accounting for atmospheric reddening at increased airmasses (i.e. \cite{fukugita1996sloan}) or using solar analog colors measured at airmasses comparable to asteroid observations may increase the accuracy of spectral slope measurements based on SDSS MOC photometry. Among the 25 asteroids with SDSS colors we note that only one asteroid predicted to have a high slope has a negative (i.e. blue) slope as observed by the LDT. This asteroid, (22689) 1998 QQ84 was observed only once in the SDSS MOC catalog, so we particularly caution against basing asteroid spectral slope measurements on single SDSS MOC observations. 

Since the SDSS MOC data do not cover wavelength ranges longer than 0.9 $\mu$m, we refrain from directly comparing spectral slopes measured by the IRTF to those predicted from the SDSS MOC data. Many asteroid taxa show a change in slope from visible to near-infrared wavelengths, particularly primitive taxa like the D-types. Additionally, highly space weathered S-types can have steep spectral slopes in the visible that do not continue into the near-IR. Therefore, we caution against extrapolating visible slopes to the near-infrared. The asteroid (31056) 1996 RK25 demonstrates that steep visible slopes do not always indicate steep near-infrared slopes. The range of visible slopes of (31056) 1996 RK25 predicted by SDSS photometry was 6.90 - 14.2 \%/0.1 $\mu$m. In the LDT data, we measured a visible spectral slope of 12.3 \%/0.1 $\mu$m, which non-uniquely indicated this object was a D-type. However, when we observed (31056) 1996 RK25 with the IRTF the prominent 1- and 2-$\mu$m absorptions in its infrared spectrum ruled out this interpretation. When the visible and near-infrared spectra were combined, the ambiguity in classification resolved, revealing (31056) 1996 RK25 is an Sw asteroid, with the w indicating a steep red slope that may be due to space weathering (\cite{demeo2009}). This example underscores the importance of including both visible and near-infrared wavelength regions in spectral classification to resolve potential ambiguity in spectral classification that cannot be addressed with visible-wavelength SDSS MOC photometry alone. 

In addition to the candidate `red' objects identified using the SDSS MOC, we also observed the asteroids (203) Pompeja and (269) Justitia following the 2021 discovery that these asteroids had extremely red, TNO-like slopes (\cite{hasegawa2021}). Our data confirm the extremely red slope of (269) Justitia, but curiously, both IRTF and LDT observations of (203) Pompeja show this object has a spectral slope more typical of an X-type asteroid. Variations in slope on (203) Pompeja have also been noted in \cite{hasegawa2022}, which suggests differences in sub-observer longitude may be responsible for spectral slope variation, with some regions of (203) Pompeja exhibiting steeply red slopes. The geometric albedo for (203) Pompeja obtained by NEOWISE is 0.036 (\cite{Mainzer_NEOWISE}), making (203) Pompeja a P-type asteroid in our observations.

\begin{figure}[!ht]

\gridline{\fig{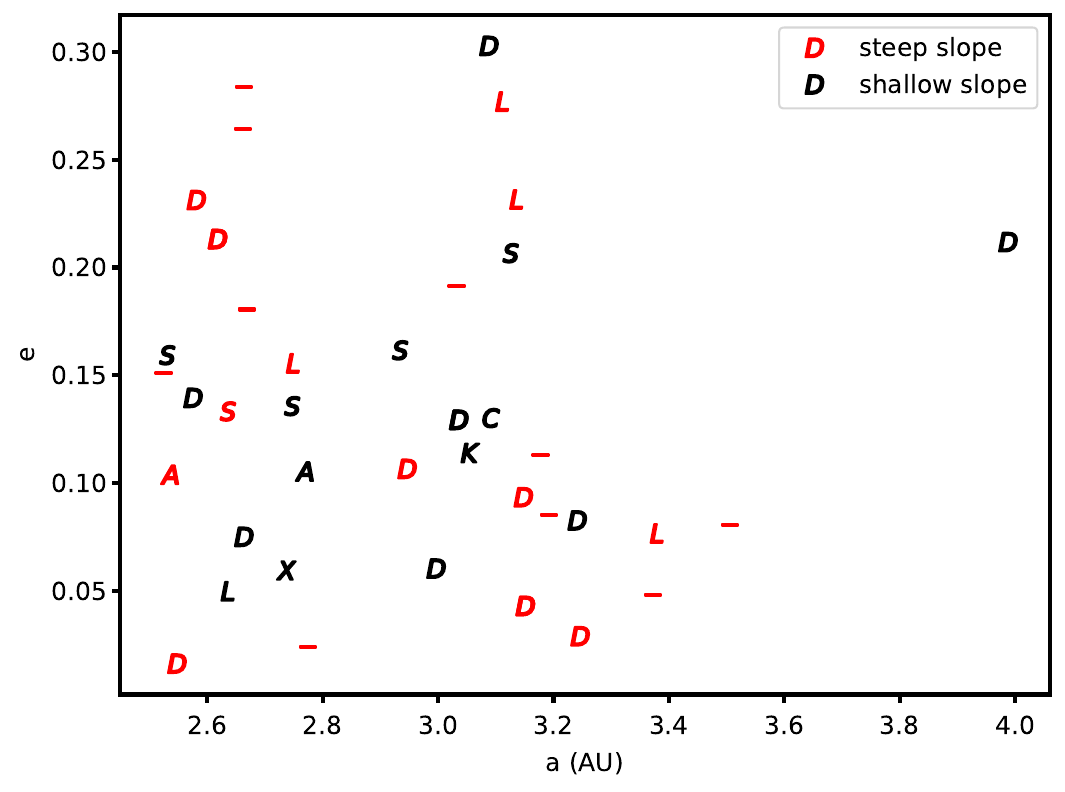}{0.5\textwidth}{(a)}
          \fig{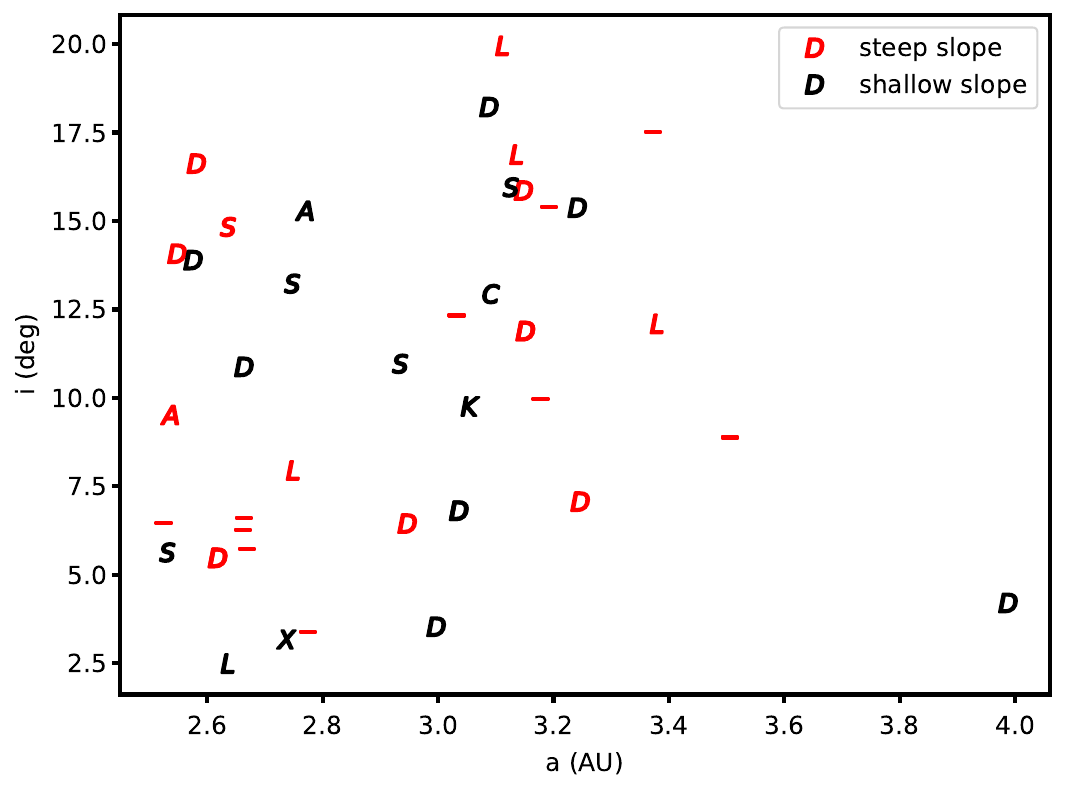}{0.5\textwidth}{(b)}
          }
\caption{Distribution of orbital elements of all the asteroids presented in this work. Panel a shows the distribution of asteroids in semimajor axis ($a$, in AU) and eccentricity ($e$) space, while Panel b shows the distribution of asteroids in semimajor axis ($a$, in AU) and inclination ($i$, in degrees) space. Asteroids with slopes exceeding the average spectral slope of the R-group Trojans in either the LDT or IRTF data are plotted in red as steep slope asteroids. Bus-DeMeo classifications for each asteroid are given by the shape of the marker. Osculating orbital elements were computed using the JPL Horizons ephemeris service for Julian Date 2460000.5 \label{fig:orb-el}}
\end{figure}

Dynamical accounts of Solar System formation posit that the distribution of asteroids in orbital element space reflects a combination of the primordial distribution of materials and later modification caused by the preferential delivery of material from different regions of the protosolar nebula to the Main Belt. In \cite{walsh2012populating}, the predominance of S-type asteroids in the inner Main Belt and C-type asteroids in the outer Main Belt are explained as a consequence of Jupiter's inward migration (in accordance with the Grand Tack model) implanting primitive asteroids from beyond the initial, compact orbit of Saturn into the Main Belt. Similarly, \cite{levison2009} demonstrates that the subsequent outward orbital migration of the giant planets into the proto-Kuiper Belt ($\sim 16 - 30$) AU resulted in the preferential delivery of primitive D- \& P-type asteroids originating from same source region as the TNOs to the outer Main Belt. In contrast to the predictions of these models, we observe a lack of observed correlations between the orbital elements $a$, $e$, and $i$ and spectral slope. We also observe a lack of correlation between the orbital elements  $a$, $e$, and $i$ and D-type classification (see Figure \ref{fig:orb-el}). This result suggests that at the relatively small asteroid sizes sampled by this survey, steep slope asteroids are thoroughly mixed throughout the Main Belt. This result is consistent with \cite{demeo2014}, which demonstrated that while a gradient in taxonomic types exists for large ($>100$ km) asteroids, this taxonomic gradient is much less pronounced for small asteroids. The detection of unexpected D-types in the inner Main Belt (\cite{demeo2014_dtype}) also supports this conclusion. This size dependence suggests that after primitive D- \& P-types were delivered to the Main Belt from their hypothesized parent population in the proto-Kuiper Belt, mechanisms of orbital evolution preferentially acted on smaller asteroids.  

We also note a diversity in near-infrared spectral shape among steeply sloped asteroids. Many steeply sloped asteroids (e.g. (27378) 2000 EG55, (81819) 2000 KS35, and (85911) 1999 CY91) are not well matched by any of the taxonomic types in the Bus-DeMeo system. Objects like these, and the steeply red-sloped asteroid (269) Justitia may be better classified using taxonomic systems developed to classify extremely red objects, such as TNO color classifications (e.g. \cite{perna2010colors}). In \cite{hasegawa2021}, (269) Justitia is classified as similar to the IR and RR TNO populations. The presence of asteroids like (269) Justitia in the Main Belt hint that the delineation between TNOs and asteroids is not entirely clear cut. Considering the asteroids and TNOs together in our evaluation of the spectra of primitive objects may improve our classification, and thus understanding, of the relationships between these liminal objects.

Some steeply sloped asteroids in our sample show a gradual shallowing of spectral slopes at long ($>1.5 \mu$m) wavelengths typical of D-type asteroids, while others show a steep and linear spectral slope that remains relatively constant over the entire near-infrared (0.7-2.5 $\mu$m) region of the spectrum. This diversity in spectral shape is further underscored by the 0.85 - J and J - K color indices of the steeply sloped asteroids. As noted in Section 4.3, there is a subpopulation of steeply sloped asteroids that have higher J - K color values than would be predicted based on their 0.85 - J color. A higher J - K color index indicates that the slopes of these asteroids do not fall off as much as expected at long wavelengths. The linear slopes of these asteroids in the near-IR distinguish them from D-type, R group Trojans as well as the RR, IR, and BR taxonomic types identified by (\cite{perna2010colors}) among Centaurs and TNOs, which all show a flattening in slope at long ($>1.5 \mu$m) wavelengths. These asteroids may represent a new, steeply red-sloped taxonomic type with spectral slopes similar to D-types from 0.75 - 1.5 $\mu$m, but higher slopes in the 1.5 - 2.5 $\mu$m region. 

To investigate the possibility that these steeply red, linearly sloped asteroids represent a distinct taxonomic classification not described in the Bus-DeMeo taxonomy, we compared the asteroids observed with the IRTF in our sample to the asteroids used to define the Bus-DeMeo taxonomic system (\cite{demeo2009}) in slope and principal component space. We directly compared the infrared slopes and infrared-only principal components of the Bus-DeMeo asteroids to the asteroids in our sample alongside the taxonomic classification of each asteroid to determine if those asteroids poorly matched to their assigned Bus-DeMeo classification  and steep red slopes (e.g. those marked with both a ``-'' and italic type in Table \ref{tab:all_res}) formed a distinct cluster in principal component space. For clarity, we refer to these asteroids which were poorly matched to their assigned Bus-DeMeo classification as ``unclassified'' asteroids. 

\begin{figure}[!ht]

\gridline{\fig{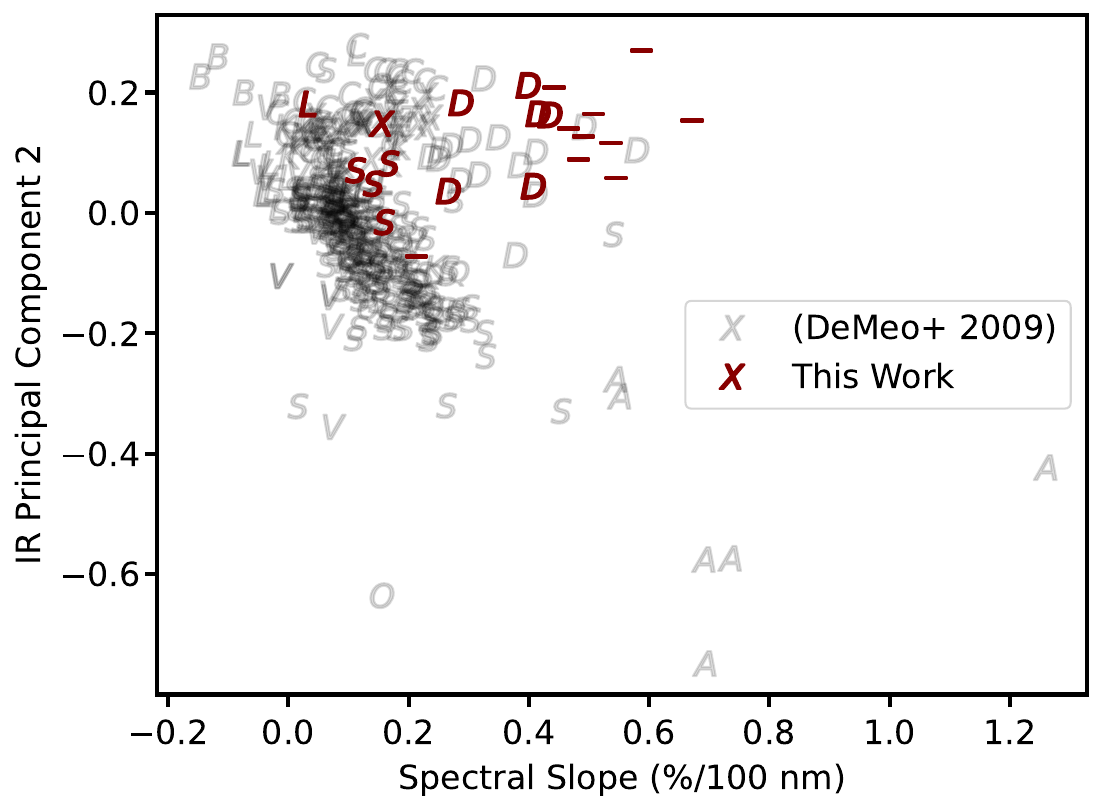}{0.33\textwidth}{(a)}
          \fig{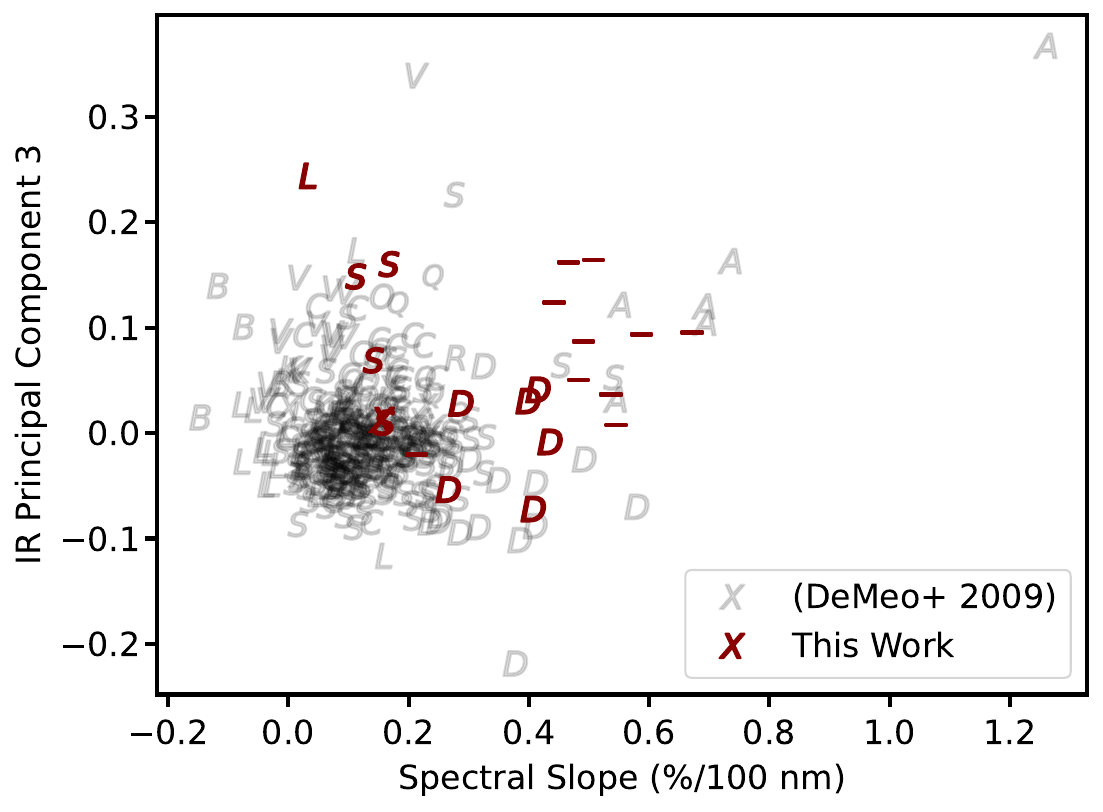}{0.33\textwidth}{(b)}
          \fig{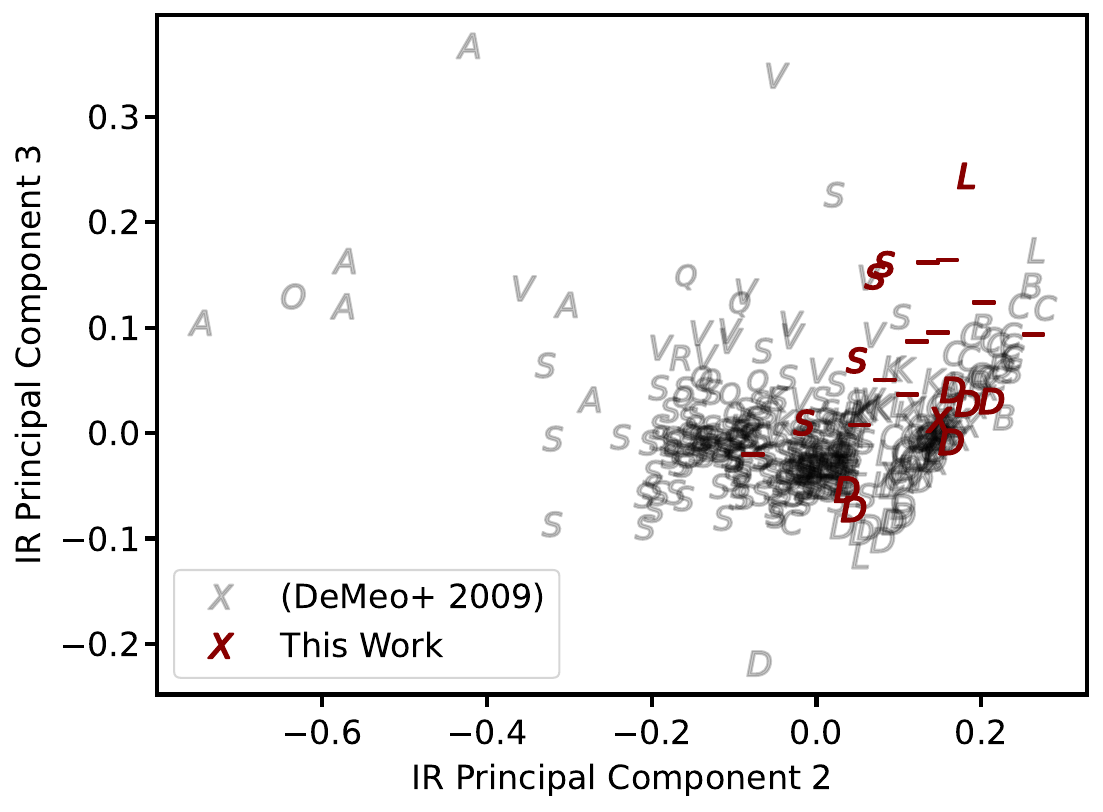}{0.33\textwidth}{(c)}
          }
\caption{Plots illustrating the clustering of steep-sloped, unclassified asteroids in spectral slope and the infrared principal component (IRPC) space defined in \cite{demeo2009}. Asteroids are plotted according to spectral slope and IRPC 2 (panel a), spectral slope and IRPC 3 (panel b), and IRPC 2 and IRPC 3 (panel c). Each asteroid's Bus-DeMeo classification is represented by the shape of the plot marker, with unclassified asteroids shown with a `-' symbol. Asteroids observed in this paper are plotted in dark red, while asteroids from the original Bus-DeMeo sample (\cite{demeo2009}) are plotted in semi-transparent grey. The clustering of the unclassified asteroids in slope and principal component space suggests that this population of steeply-sloped asteroids may represent a distinct taxonomic class not described in the Bus-DeMeo taxonomic system. \label{fig:PCA}}
\end{figure}

In terms of slope, the unclassified asteroids tend to have steeper slopes than the typical D-type asteroid. All unclassified asteroids in our sample except for (7562) Kagiroino-Oka have infrared spectral slopes exceeding those of the typical R-group Trojan, though not all asteroids with slope exceeding the typical R-group Trojan are unclassified (see Table \ref{tab:all_res}). The ranges of slopes among D-types and the unclassified asteroids overlap. Similarly, the range of slopes of the steeply red sloped, unclassified asteroids overlaps with other populations within the Bus-DeMeo taxonomy including the A-types and some highly space-weathered S-types (see Figure \ref{fig:PCA}, panel a). Slope alone then, does not distinguish the unclassified asteroids from established taxonomic types, but the values of other principal components may.

We identify principal components 2 and 3 (IRPC 2 and IRPC 3) as potential means to distinguish between these different classifications of high slope asteroids. We suggest using IRPC 2 to distinguish between A-types and the unclassified asteroids as the A-types are characterized by negative values of IRPC 2 while the unclassified asteroids are characterized by positive values of IRPC 2. The space weathered S-types may also be distinguished in IRPC 2 vs. slope space, as the more steeply sloped S-types tend to have lower values of IRPC 2 than the unclassified asteroids. In IRPC 2 vs. slope space (Figure \ref{fig:PCA}, panel a), the unclassified asteroids still plot within the D-type envelope. Therefore, additional principal components need to be used to distinguish between the D-types and unclassified asteroids. In particular, we find that the unclassified asteroids tend to have a higher than average values of IRPC 3 than the D-types, though these two populations again overlap in range. The unclassified asteroids can be distinguished from the D-type population using a combination of IRPC 3 and spectral slope. In IRPC 3 vs. slope space (Figure \ref{fig:PCA}, panel b), the unclassified asteroids still plot within the S-type and A-type envelopes, though as they can be distinguished using IRPC 2. Finally, while the unclassified asteroids overlap with the S-, C-, K-, and L-type asteroids in IRPC 2 vs. IRPC 3 space (Figure \ref{fig:PCA}, panel c), the unclassified asteroids can be distinguished from these types via their steep spectral slopes. 

While these trends are suggestive, the relative rarity of steeply sloped Main Belt asteroids (especially compared to the more robust populations of S- and C-complex asteroids in \cite{demeo2009}) and smaller sample size of this survey makes drawing definite conclusions about the population of unclassified asteroids difficult. In reality, the distinction between steeply-sloped, unclassified asteroids and the taxonomic types defined by in \cite{demeo2009} in principal component space may be blurrier than the delineations suggested here. Identifying and characterizing additional steeply-sloped asteroids will help to better define the shape of the cluster tentatively identified here in principal component space. 

In \cite{mahlke2022}, the Bus-DeMeo classification scheme is modified to allow for the classification of visible-only data and re-introduces albedo as a possible classification criterion by using probabilistic classification based on available observables. That paper also introduces a new taxonomic class, the Z-types, which are characterized by steep red slopes (\cite{mahlke2022}), similar to the unclassified asteroids identified in this paper. We applied the Mahlke et. al classification scheme, including reported albedo from \cite{Mainzer_NEOWISE} when available, to the asteroid spectra presented here. We find that, among the unclassified asteroids,
\begin{itemize} 
\item six are re-classified as Z-types: (22110) 2000 QR7, (27378) 2000 EG55, (67244) 2000 EH58, (76391) 2000 FP7, (81819) 2000 KS35, and (85911) 1999 CY91,
\item three are re-classified as D-types: (21867) 1999 TQ251, (22422) Kenmount Hill, and (25835) Tomzega,
\item one is re-classified as an S-type: (1947) Iso-Heikkila.
\end{itemize} 
The reclassification of a portion of the unclassified asteroids as Z-types supports the existence of a taxonomic class not present in the Bus-DeMeo system and tentatively identifies this new population with the Z-types of \cite{mahlke2022}. 

\begin{figure}
    \centering
    \includegraphics[width=0.75\linewidth]{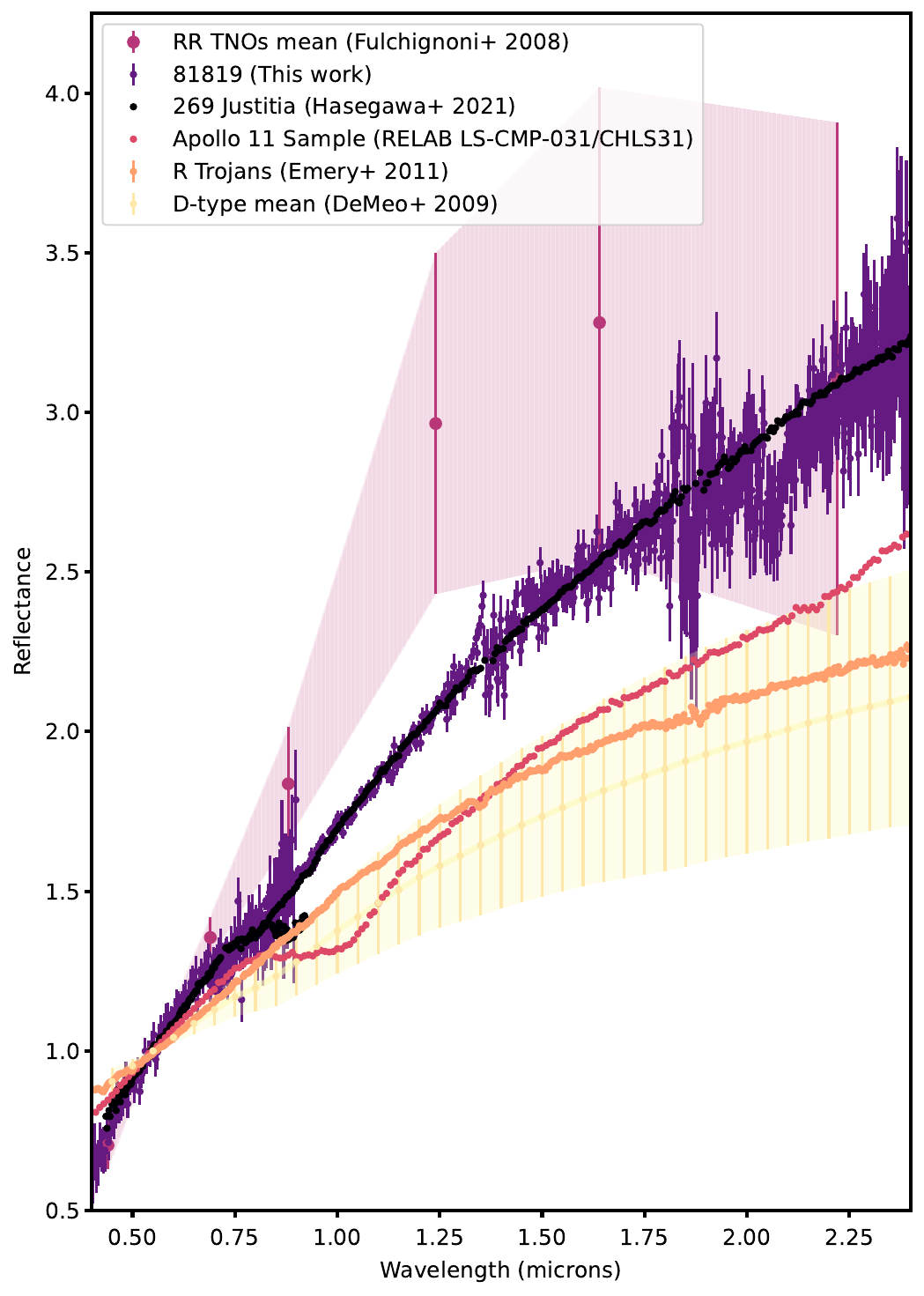}
    \caption{Comparison of the visible and near-infrared spectra of objects throughout the Solar System with steeply red spectral slopes. The VNIR spectrum of steep sloped red object (81819) 2000 KS35 from this work is plotted alongside examples from the literature for comparison purposes. The mean colors of the RR TNOs and accompanying 1-$\sigma$ color ranges are sourced from \cite{fulchignoni2008transneptunian}. The spectrum of large, steeply-sloped Main Belt asteroid 269 Justitia originates from \cite{hasegawa2021}. The mean R Trojan spectra is reproduced from \cite{Emery_2010}. The Apollo 11 sample measurements are taken from the RELAB database (\cite{PDS_RELAB}) and were originally measured by Carle M. Pieters. The D-type average spectrum and 1-$\sigma$ ranges are sourced from \cite{demeo2009}. All spectra are normalized to 1 at 0.55 $\mu$m for ease of comparison.}
    \label{fig:bered}
\end{figure}

If the unclassified (or Z-type) subpopulation represents a newly-identified population of red-sloped asteroids, its existence could challenge the assumption that all steeply red sloped asteroids originate in the outer Solar System. The steep red slopes of Centaurs and TNOs at short wavelengths ($<1.5\mu$m) and subsequent flattening at longer wavelengths ($>1.5 \mu$m) are well matched by organic compounds irradiated in the lab and are consistent with the chemical species predicted to be present in the TNO region (\cite{hudson2008laboratory, brunetto2006ion}). Similarly, the observed colors of the Jupiter Trojans can be explained by the initial formation of a TNO-like, organic-rich irradiation crust in the proto-Kuiper Belt followed by a neutralization in slope caused by a loss of volatile ices during the Trojan's migration to their current locations closer to the Sun (\cite{Wong_2016}). In contrast, the consistently linear slopes of the unclassified asteroids are not well matched by irradiated organics. Consistently linear, steeply-red slopes with weakened absorption features are instead also found on space weathered surfaces in the inner Solar System, including mature, heavily irradiated lunar soils (\cite{adams1970spectral, gaffey2010space}) and the moons of Mars (\cite{MURCHIE199663, rivkin2002phobos}). The resemblance of the unclassified and Z-type asteroids to the space weathered surfaces of the moons of Earth and Mars challenges the interpretation that these asteroids are primitive asteroids originating from beyond the orbit of Jupiter. Lunar soils, for example, are not primitive as they have undergone chemical processing and differentiation within the early lunar magma ocean \cite{wieczorek2006constitution}. While the origins of the martian moons Phobos and Deimos are still debated, some authors suggest these moons formed in-situ in a circum-Martian debris disk, while others posit they are captured D-type asteroids based on their resemblance to Trojans and other Main Belt D-types (\cite{rosenblatt2011origin}). Comparison of the spectra of Phobos and Deimos to the Z-type and other steeply-sloped asteroids may shed light on the origins of these enigmatic moons. 

Examples of the varied VNIR spectral appearances of steeply-red sloped materials throughout the Solar System are illustrated in Figure \ref{fig:bered}. Steeply red slopes are seen both on objects thought to be primitive objects thought to originate in the outer Solar System including the RR TNOs (\cite{fulchignoni2008transneptunian}) and R Trojans (\cite{Emery_2010}). Red slopes are also found on materials that have undergone significant processing, as in the Apollo 11 sample of bulk lunar regolith (\cite{PDS_RELAB}, see also \cite{adams1970spectral}). The steeply red-sloped subpopulation discussed in this work is represented by (81819) 2000 KS35 and most closely matches observations of large Main Belt asteroid (269) Justitia by \cite{hasegawa2021}. These two asteroids differ from the average D-type asteroid in \cite{demeo2009} and the R group Trojans with their significantly steeper slopes and relatively constant slopes over the 0.4 - 2.4 $\mu$m range. While the average slope of the RR TNOs over this range is comparable to the slopes of (269) Justitia and (81819) 2000 KS35, the shape of the spectrum of the average RR TNO also lacks the linear shape seen in the spectra of those two Main Belt asteroids. While the lunar sample has a steep, linear red slope over the 0.4-2.4 $\mu$m range, the absorption near 1 $\mu$m is not present in the asteroid spectra. The differences between the spectra of (269) Justitia and (81819) 2000 KS35 and the other small body populations presented suggest that these asteroid may represent a distinct subpopulation of steeply-sloped asteroids in the Main Belt.

The existence of a newly-identified population of steeply-sloped asteroids and their possible inner (e.g. interior to Jupiter) Solar System origins agrees with the results of \cite{gartrelle2021_familyneighborhoods}, which explores the subtle differences in spectral appearances of D-type asteroids at different heliocentric distances. That work found that Trojans tended to have lower slopes at long wavelengths than Main Belt asteroids. Similarly, D-type asteroids at lower heliocentric distances had steeper slopes in the 1.5–1.8 $\mu$m and 2.0–2.45 $\mu$m segments than those at higher heliocentric distances (\cite{gartrelle2021_familyneighborhoods}). The increased prevalence of asteroids with steep slopes in the 1.5-2.45 $\mu$m region could indicate that these asteroids are sourced from regions interior to the orbit of Jupiter. Alternatively, increasing 1.5-2.45 $\mu$m slopes at lower heliocentric distances may indicate steepening slopes are a product of exposing primitive, TNO-like material to the increased thermal and ionizing radiation environment closer to the Sun. Laboratory studies of how organic irradiation crusts evolve spectrally when exposed to the thermal environment of the Main Belt could provide support for the idea that asteroids with linearly increasing, steep red slopes share a parent population with the typical D-type asteroids in the outer Main Belt and Trojan population. Particularly, if laboratory work demonstrates that the loss of volatile compounds at the higher temperatures expected in the inner Main Belt result in a steepening in the 1.5-2.45 $\mu$m spectral region, the subtle differences in slope between D-type and other steeply red-sloped asteroids may be explained by differences in thermal histories. 

\section{Conclusions}

Using the Sloan Digital Sky Survey's Moving Object Catalog, we identified 60 `red' candidate objects with spectrophotometric slopes exceeding the average slope of the R group Trojans. Using the NASA IRTF and the Lowell Discovery Telescope we conducted a survey to determine the spectral slopes and taxonomic classifications of 30 of these candidate objects to verify their steep spectral slopes and compare their spectra to those of Jupiter Trojans and other steeply red Solar System objects. We find that using our selection criteria, $\sim 50\%$ of candidate `red' objects in the SDSS MOC have slopes equal to or exceeding the slope of the average R group Trojan. Examining the distribution of orbital elements of these objects, we find no obvious correlations between semimajor axis ($a$), eccentricity ($e$), and inclination ($i$) and spectral slope, indicating steeply red sloped objects are well mixed within the Main Belt. We additionally identify a diversity in spectral appearance among steeply sloped asteroids, which suggests multiple sub-types of red-sloped asteroids are present within the Main Belt. This spectral diversity among red-sloped asteroids hints at a multiplicity of origins for spectrally red-sloped material within the Main Belt. Further explorations of the visible and near-infrared spectra of these objects and integration of classification methods developed for asteroids with those developed for trans-Neptunian objects and Centaurs will lead to a greater understanding of the population of unusually red-sloped Main Belt asteroids.

\section{Acknowledgements}

The authors would like to thank Josh Emery, Will Grundy, and Chad Trujillo for their thoughtful comments which greatly influenced the direction of discussion section. The authors also thank Francesca DeMeo for providing the principal component values for the asteroids used to define the Bus-DeMeo taxonomy. 

This material is based upon work supported by the National Aeronautics and Space Administration under Grant No. 80NSSC20K0671 issued through Science Mission Directorate via the ROSES-2019 Solar System Observations program. 

These results made use of the Lowell Discovery Telescope (LDT) at Lowell Observatory. Lowell is a private, non-profit institution dedicated to astrophysical research and public appreciation of astronomy and operates the LDT in partnership with Boston University, the University of Maryland, the University of Toledo, Northern Arizona University and Yale University. The upgrade of the DeVeny optical spectrograph has been funded by a generous grant from John and Ginger Giovale and by a grant from the Mt. Cuba Astronomical Foundation. Lowell observatory sits at the base of the San Francisco Peaks, on homelands sacred to Native Americans throughout the region. We honor their past, present, and future generations, who have lived here for millennia and will forever call this place home.

The authors acted as visiting Astronomers at the Infrared Telescope Facility, which is operated by the University of Hawaii under contract 80HQTR19D0030 with the National Aeronautics and Space Administration. The authors wish to recognize and acknowledge the very significant cultural role and reverence that the summit of Maunakea has always had within the indigenous Hawaiian community.  We are most fortunate to have the opportunity to conduct observations from this mountain. 

This research has made use of data and/or services provided by the International Astronomical Union's Minor Planet Center. This research has made use of the SIMBAD database, operated at CDS, Strasbourg, France. Taxonomic type results presented in this work were determined, in whole or in part, using the Bus-DeMeo Taxonomy Classification Web tool by Stephen M. Slivan, developed at MIT with the support of National Science Foundation Grant 0506716 and NASA Grant NAG5-12355. This research also utilizes spectra acquired by Carle M. Pieters with the NASA RELAB facility at Brown University.

\bibliography{sample631}{}

\begin{thebibliography}{}
\expandafter\ifx\csname natexlab\endcsname\relax\def\natexlab#1{#1}\fi
\providecommand{\url}[1]{\href{#1}{#1}}
\providecommand{\dodoi}[1]{doi:~\href{http://doi.org/#1}{\nolinkurl{#1}}}
\providecommand{\doeprint}[1]{\href{http://ascl.net/#1}{\nolinkurl{http://ascl.net/#1}}}
\providecommand{\doarXiv}[1]{\href{https://arxiv.org/abs/#1}{\nolinkurl{https://arxiv.org/abs/#1}}}

\bibitem[{Adams \& Jones(1970)}]{adams1970spectral}
Adams, J.~B., \& Jones, R.~L. 1970, Science, 167, 737

\bibitem[{{Bida} {et~al.}(2014){Bida}, {Dunham}, {Massey}, \& {Roe}}]{DeVeny}
{Bida}, T.~A., {Dunham}, E.~W., {Massey}, P., \& {Roe}, H.~G. 2014, in Society
  of Photo-Optical Instrumentation Engineers (SPIE) Conference Series, Vol.
  9147, Ground-based and Airborne Instrumentation for Astronomy V, ed. S.~K.
  {Ramsay}, I.~S. {McLean}, \& H.~{Takami}, 91472N, \dodoi{10.1117/12.2056872}

\bibitem[{Bottke~Jr {et~al.}(2006)Bottke~Jr, Vokrouhlick{\`y}, Rubincam, \&
  Nesvorn{\`y}}]{bottke2006yarkovsky}
Bottke~Jr, W.~F., Vokrouhlick{\`y}, D., Rubincam, D.~P., \& Nesvorn{\`y}, D.
  2006, Annu. Rev. Earth Planet. Sci., 34, 157,
  \dodoi{https://doi.org/10.1146/annurev.earth.34.031405.125154}

\bibitem[{Brown {et~al.}(2011)Brown, Schaller, \& Fraser}]{brown2011hypothesis}
Brown, M., Schaller, E., \& Fraser, W. 2011, The Astrophysical Journal Letters,
  739, L60, \dodoi{10.1088/2041-8205/739/2/L60}

\bibitem[{Brunetto {et~al.}(2006)Brunetto, Barucci, Dotto, \&
  Strazzulla}]{brunetto2006ion}
Brunetto, R., Barucci, M.~A., Dotto, E., \& Strazzulla, G. 2006, The
  Astrophysical Journal, 644, 646, \dodoi{10.1086/503359}

\bibitem[{Bus {et~al.}(2011)Bus, Gulbis, Elliot, Denault, Rayner, Stahlberger,
  Chung, \& Tokunaga}]{bus2011moris}
Bus, S., Gulbis, A., Elliot, J., {et~al.} 2011, in EPSC-DPS Joint Meeting, 1834

\bibitem[{Bus(1999)}]{bus1999}
Bus, S.~J. 1999, Ph. D. Thesis, 311

\bibitem[{Bus \& Binzel(2002)}]{bus2002}
Bus, S.~J., \& Binzel, R.~P. 2002, Icarus, 158, 106,
  \dodoi{https://doi.org/10.1006/icar.2002.6857}

\bibitem[{Casagrande {et~al.}(2012)Casagrande, Ram{\'\i}rez, Melendez, \&
  Asplund}]{casagrande2012infrared}
Casagrande, L., Ram{\'\i}rez, I., Melendez, J., \& Asplund, M. 2012, The
  Astrophysical Journal, 761, 16, \dodoi{10.1088/0004-637X/761/1/16}

\bibitem[{Cushing {et~al.}(2004)Cushing, Vacca, \&
  Rayner}]{cushing2004spextool}
Cushing, M.~C., Vacca, W.~D., \& Rayner, J.~T. 2004, Publications of the
  Astronomical Society of the Pacific, 116, 362, \dodoi{10.1086/382907}

\bibitem[{Cutri {et~al.}(2003)Cutri, Skrutskie, Van~Dyk, Beichman, Carpenter,
  Chester, Cambresy, Evans, Fowler, Gizis, {et~al.}}]{cutri20032mass}
Cutri, R., Skrutskie, M., Van~Dyk, S., {et~al.} 2003, The IRSA 2MASS All-Sky
  Point Source Catalog

\bibitem[{DeMeo \& Carry(2013)}]{demeo2014}
DeMeo, F., \& Carry, B. 2013, Icarus, 226, 723,
  \dodoi{https://doi.org/10.1016/j.icarus.2013.06.027}

\bibitem[{DeMeo {et~al.}(2014)DeMeo, Binzel, Carry, Polishook, \&
  Moskovitz}]{demeo2014_dtype}
DeMeo, F.~E., Binzel, R.~P., Carry, B., Polishook, D., \& Moskovitz, N.~A.
  2014, Icarus, 229, 392, \dodoi{https://doi.org/10.1016/j.icarus.2013.11.026}

\bibitem[{DeMeo {et~al.}(2009)DeMeo, Binzel, Slivan, \& Bus}]{demeo2009}
DeMeo, F.~E., Binzel, R.~P., Slivan, S.~M., \& Bus, S.~J. 2009, Icarus, 202,
  160

\bibitem[{DeMeo \& Carry(2014)}]{DemeoCarry_2014}
DeMeo, F.~E., \& Carry, B. 2014, Nature, 505, 629,
  \dodoi{https://doi.org/10.1038/nature12908}

\bibitem[{{Devogele} \& {Moskovitz}(2019)}]{Devogele_SP}
{Devogele}, M., \& {Moskovitz}, N. 2019, in EPSC-DPS Joint Meeting 2019, Vol.
  2019, EPSC--DPS2019--841

\bibitem[{Doressoundiram {et~al.}(2008)Doressoundiram, Boehnhardt, Tegler, \&
  Trujillo}]{doressoundiram2008color}
Doressoundiram, A., Boehnhardt, H., Tegler, S.~C., \& Trujillo, C. 2008, The
  solar system beyond Neptune, 91

\bibitem[{Emery {et~al.}(2011)Emery, Burr, \& Cruikshank}]{Emery_2010}
Emery, J.~P., Burr, D.~M., \& Cruikshank, D.~P. 2011, The Astronomical Journal,
  141, 25, \dodoi{https://doi.org/10.1088/0004-6256/141/1/25}

\bibitem[{Fukugita {et~al.}(1996)Fukugita, Ichikawa, Gunn, Doi, Shimasaku, \&
  Schneider}]{fukugita1996sloan}
Fukugita, M., Ichikawa, T., Gunn, J., {et~al.} 1996, Astronomical Journal v.
  111, p. 1748, 111, 1748

\bibitem[{Fulchignoni {et~al.}(2008)Fulchignoni, Belskaya, Barucci, De~Sanctis,
  \& Doressoundiram}]{fulchignoni2008transneptunian}
Fulchignoni, M., Belskaya, I., Barucci, M.~A., De~Sanctis, M.~C., \&
  Doressoundiram, A. 2008, The Solar System Beyond Neptune, 181

\bibitem[{Gaffey(2010)}]{gaffey2010space}
Gaffey, M.~J. 2010, Icarus, 209, 564,
  \dodoi{https://doi.org/10.1016/j.icarus.2010.05.006}

\bibitem[{Gartrelle {et~al.}(2021{\natexlab{a}})Gartrelle, Hardersen, Izawa, \&
  Nowinski}]{gartrelle_2021}
Gartrelle, G.~M., Hardersen, P.~S., Izawa, M.~R., \& Nowinski, M.~C.
  2021{\natexlab{a}}, Icarus, 354, 114043,
  \dodoi{https://doi.org/10.1016/j.icarus.2020.114043}

\bibitem[{Gartrelle {et~al.}(2021{\natexlab{b}})Gartrelle, Hardersen, Izawa, \&
  Nowinski}]{gartrelle2021_familyneighborhoods}
---. 2021{\natexlab{b}}, Icarus, 363, 114295

\bibitem[{Gradie \& Tedesco(1982)}]{gradie1982}
Gradie, J., \& Tedesco, E. 1982, Science, 216, 1405

\bibitem[{Gulbis {et~al.}(2010)Gulbis, Elliot, Rojas, Bus, Rayner, Stahlberger,
  Tokunaga, Adams, \& Person}]{gulbis2010moris}
Gulbis, A.~A., Elliot, J., Rojas, F., {et~al.} 2010, in AAS/Division for
  Planetary Sciences Meeting Abstracts\# 42, Vol.~42, 49--14

\bibitem[{Hainaut {et~al.}(2012)Hainaut, Boehnhardt, \&
  Protopapa}]{hainaut2012colours}
Hainaut, O., Boehnhardt, H., \& Protopapa, S. 2012, Astronomy \& Astrophysics,
  546, A115, \dodoi{https://doi.org/10.1051/0004-6361/201219566}

\bibitem[{Hasegawa {et~al.}(2021)Hasegawa, Marsset, DeMeo, Bus, Geem, Ishiguro,
  Im, Kuroda, \& Vernazza}]{hasegawa2021}
Hasegawa, S., Marsset, M., DeMeo, F.~E., {et~al.} 2021, The Astrophysical
  Journal Letters, 916, L6, \dodoi{10.3847/2041-8213/ac0f05}

\bibitem[{Hasegawa {et~al.}(2022)Hasegawa, DeMeo, Marsset, Hanu{\v{s}},
  Avdellidou, Delbo, Bus, Hanayama, Horiuchi, Takir, {et~al.}}]{hasegawa2022}
Hasegawa, S., DeMeo, F.~E., Marsset, M., {et~al.} 2022, The Astrophysical
  journal letters, 939, L9, \dodoi{10.3847/2041-8213/ac92e4}

\bibitem[{Hudson {et~al.}(2008)Hudson, Palumbo, Strazzulla, Moore, Cooper, \&
  Sturner}]{hudson2008laboratory}
Hudson, R., Palumbo, M., Strazzulla, G., {et~al.} 2008, The Solar System Beyond
  Neptune, 507

\bibitem[{Ivezi{\'c} {et~al.}(2002)Ivezi{\'c}, Juric, Lupton, Tabachnik, \&
  Quinn}]{ivezic2002asteroids}
Ivezi{\'c}, {\v{Z}}., Juric, M., Lupton, R.~H., Tabachnik, S., \& Quinn, T.
  2002, in Survey and Other Telescope Technologies and Discoveries, Vol. 4836,
  SPIE, 98--103, \dodoi{https://doi.org/10.1117/12.457304}

\bibitem[{Ivezi{\'c} {et~al.}(2001)Ivezi{\'c}, Tabachnik, Rafikov, Lupton,
  Quinn, Hammergren, Eyer, Chu, Armstrong, Fan, {et~al.}}]{ivezic2001solar}
Ivezi{\'c}, {\v{Z}}., Tabachnik, S., Rafikov, R., {et~al.} 2001, The
  Astronomical Journal, 122, 2749, \dodoi{10.1086/323452}

\bibitem[{Lamy \& Toth(2009)}]{lamy2009colors}
Lamy, P., \& Toth, I. 2009, Icarus, 201, 674,
  \dodoi{https://doi.org/10.1016/j.icarus.2009.01.030}

\bibitem[{Levison {et~al.}(2009)Levison, Bottke, Gounelle, Morbidelli,
  Nesvorn{\`y}, \& Tsiganis}]{levison2009}
Levison, H.~F., Bottke, W.~F., Gounelle, M., {et~al.} 2009, Nature, 460, 364,
  \dodoi{https://doi.org/10.1038/nature08094}

\bibitem[{Mahlke {et~al.}(2022)Mahlke, Carry, \& Mattei}]{mahlke2022}
Mahlke, M., Carry, B., \& Mattei, P.-A. 2022, Astronomy \& Astrophysics, 665,
  A26, \dodoi{https://doi.org/10.1051/0004-6361/202243587}

\bibitem[{Mainzer {et~al.}(2019)Mainzer, Bauer, Cutri, Grav, Kramer, Masiero, ,
  Wright, \& Eds.}]{Mainzer_NEOWISE}
Mainzer, A., Bauer, J., Cutri, R., {et~al.} 2019, NEOWISE Diameters and Albedos
  V2.0,  NASA Planetary Data System, \dodoi{10.26033/18s3-2z54}

\bibitem[{Marsset {et~al.}(2020)Marsset, DeMeo, Binzel, Bus, Burbine, Burt,
  Moskovitz, Polishook, Rivkin, Slivan, {et~al.}}]{marsset2020}
Marsset, M., DeMeo, F.~E., Binzel, R.~P., {et~al.} 2020, The Astrophysical
  Journal Supplement Series, 247, 73,
  \dodoi{https://doi.org/10.3847/1538-4365/ab7b5f}

\bibitem[{Milliken(2020)}]{PDS_RELAB}
Milliken, R. 2020, RELAB Spectral Library Bundle,  NASA Planetary Data System,
  \dodoi{https://doi.org/10.17189/1519032}

\bibitem[{Morbidelli {et~al.}(2005)Morbidelli, Levison, Tsiganis, \&
  Gomes}]{Morbidelli_2005}
Morbidelli, A., Levison, H.~F., Tsiganis, K., \& Gomes, R. 2005, Nature, 435,
  462, \dodoi{https://doi.org/10.1038/nature03540}

\bibitem[{Murchie \& Erard(1996)}]{MURCHIE199663}
Murchie, S., \& Erard, S. 1996, Icarus, 123, 63,
  \dodoi{https://doi.org/10.1006/icar.1996.0142}

\bibitem[{Nesvorn{\`y} {et~al.}(2013)Nesvorn{\`y}, Vokrouhlick{\`y}, \&
  Morbidelli}]{Nesvorny_2013}
Nesvorn{\`y}, D., Vokrouhlick{\`y}, D., \& Morbidelli, A. 2013, The
  Astrophysical Journal, 768, 45,
  \dodoi{https://doi.org/10.1088/0004-637X/768/1/45}

\bibitem[{Perna {et~al.}(2010)Perna, Barucci, Fornasier, Demeo, Alvarez-Candal,
  Merlin, Dotto, Doressoundiram, \& de~Bergh}]{perna2010colors}
Perna, D., Barucci, M.~A., Fornasier, S., {et~al.} 2010, Astronomy \&
  Astrophysics, 510, A53, \dodoi{https://doi.org/10.1051/0004-6361/200913654}

\bibitem[{Pirani {et~al.}(2019)Pirani, Johansen, Bitsch, Mustill, \&
  Turrini}]{pirani_2019}
Pirani, S., Johansen, A., Bitsch, B., Mustill, A.~J., \& Turrini, D. 2019,
  Astronomy \& Astrophysics, 623, A169

\bibitem[{Prochaska {et~al.}(2020)Prochaska, Hennawi, Westfall, Cooke, Wang,
  Hsyu, Davies, Farina, \& Pelliccia}]{pypeit:joss_pub}
Prochaska, J.~X., Hennawi, J.~F., Westfall, K.~B., {et~al.} 2020, Journal of
  Open Source Software, 5, 2308, \dodoi{10.21105/joss.02308}

\bibitem[{{Prochaska} {et~al.}(2020){Prochaska}, {Hennawi}, {Cooke},
  {Westfall}, {Wang}, {EmAstro}, {Tiffanyhsyu}, {Wasserman}, {Villaume},
  {Marijana777}, {Schindler}, {Young}, {Simha}, {Wilde}, {Tejos}, {Isbell},
  {Fl{\"o}rs}, {Sandford}, {Vasovi{\'c}}, {Betts}, \& {Holden}}]{pypeit:zenodo}
{Prochaska}, J.~X., {Hennawi}, J., {Cooke}, R., {et~al.} 2020, {pypeit/PypeIt:
  Release 1.0.0}, v1.0.0,  Zenodo, \dodoi{10.5281/zenodo.3743493}

\bibitem[{Rayner {et~al.}(2003)Rayner, Toomey, Onaka, Denault, Stahlberger,
  Vacca, Cushing, \& Wang}]{rayner2003spex}
Rayner, J.~T., Toomey, D., Onaka, P., {et~al.} 2003, Publications of the
  Astronomical Society of the Pacific, 115, 362, \dodoi{10.1086/367745}

\bibitem[{Rivkin {et~al.}(2002)Rivkin, Brown, Trilling, Bell~Iii, \&
  Plassmann}]{rivkin2002phobos}
Rivkin, A., Brown, R., Trilling, D., Bell~Iii, J., \& Plassmann, J. 2002,
  Icarus, 156, 64, \dodoi{https://doi.org/10.1006/icar.2001.6767}

\bibitem[{Rosenblatt(2011)}]{rosenblatt2011origin}
Rosenblatt, P. 2011, Astronomy \& Astrophysics Review, 19,
  \dodoi{10.1007/s00159-011-0044-6}

\bibitem[{Sheppard(2010)}]{sheppard2010colors}
Sheppard, S.~S. 2010, The Astronomical Journal, 139, 1394,
  \dodoi{10.1088/0004-6256/139/4/1394}

\bibitem[{Souza-Feliciano {et~al.}(2020)Souza-Feliciano, {De Prá},
  Pinilla-Alonso, Alvarez-Candal, Fernández-Valenzuela, {De León}, Binzel,
  Arcoverde, Rondón, \& Evangelista}]{SouzaFeliciano_2020}
Souza-Feliciano, A., {De Prá}, M., Pinilla-Alonso, N., {et~al.} 2020, Icarus,
  338, 113463, \dodoi{https://doi.org/10.1016/j.icarus.2019.113463}

\bibitem[{Walsh {et~al.}(2012)Walsh, Morbidelli, Raymond, O’brien, \&
  Mandell}]{walsh2012populating}
Walsh, K.~J., Morbidelli, A., Raymond, S.~N., O’brien, D., \& Mandell, A.
  2012, Meteoritics \& Planetary Science, 47, 1941,
  \dodoi{https://doi.org/10.1111/j.1945-5100.2012.01418.x}

\bibitem[{Wenger {et~al.}(2000)Wenger, Ochsenbein, Egret, Dubois, Bonnarel,
  Borde, Genova, Jasniewicz, Lalo{\"e}, Lesteven, {et~al.}}]{simbad}
Wenger, M., Ochsenbein, F., Egret, D., {et~al.} 2000, Astronomy and
  Astrophysics Supplement Series, 143, 9,
  \dodoi{https://doi.org/10.1051/aas:2000332}

\bibitem[{Wieczorek {et~al.}(2006)Wieczorek, Jolliff, Khan, Pritchard, Weiss,
  Williams, Hood, Righter, Neal, Shearer, {et~al.}}]{wieczorek2006constitution}
Wieczorek, M.~A., Jolliff, B.~L., Khan, A., {et~al.} 2006, Reviews in
  mineralogy and geochemistry, 60, 221,
  \dodoi{https://doi.org/10.2138/rmg.2006.60.3}

\bibitem[{Wong \& Brown(2016)}]{Wong_2016}
Wong, I., \& Brown, M.~E. 2016, The Astronomical Journal, 152, 90,
  \dodoi{https://doi.org/10.3847/0004-6256/152/4/90}

\bibitem[{Wong {et~al.}(2014)Wong, Brown, \& Emery}]{Wong_2014}
Wong, I., Brown, M.~E., \& Emery, J.~P. 2014, The Astronomical Journal, 148,
  112, \dodoi{https://doi.org/10.1088/0004-6256/148/6/112}

\end{thebibliography}
\bibliographystyle{aasjournal}

\end{document}